\documentclass[journal]{IEEEtran}

\IEEEoverridecommandlockouts

\usepackage[cmex10]{amsmath}
\usepackage{amssymb}
\usepackage{graphicx}
\usepackage{subfigure}
\usepackage{cite}
\usepackage{epstopdf}
\usepackage{subeqnarray}
\usepackage{diagbox}
\usepackage{booktabs}
\usepackage{multirow}
\usepackage{multicol}
\usepackage{algorithm}
\usepackage{algorithmic}
\usepackage{url}
\usepackage{xcolor}
\usepackage{array}
\usepackage{footnote}
\usepackage{enumerate}
\usepackage{framed}
\usepackage{lineno}
\usepackage{threeparttable}
\makesavenoteenv{tabular}

\newcommand{\mb}{\mathbf}

\newcommand{\mc}{\mathcal}

\newtheorem{theorem}{\textbf{Theorem}}
\newtheorem{lemma}{\textbf{Lemma}}

\newtheorem{corollary}{\textbf{Corollary}}
\newtheorem{definition}{\textbf{Definition}}

\newtheorem{conjecture}{\textbf{Conjecture}}
\newtheorem{construction}{\textbf{Construction}}

\begin{document}

\title{A Class of Optimal Structures for Node Computations in Message Passing Algorithms}

\author{%
\IEEEauthorblockN{Xuan He, Kui Cai, and Liang Zhou}
\thanks{
Xuan He is with the School of Information Science and
Technology, Southwest Jiaotong University, Chengdu 611756, China (e-mail:
xhe@swjtu.edu.cn).}
\thanks{
Kui Cai is with the Science, Mathematics and
Technology (SMT) Cluster, Singapore University of Technology and Design,
Singapore 487372 (e-mail: cai\_kui@sutd.edu.sg).}
\thanks{
Liang Zhou is with the National Key Laboratory of Science and Technology
on Communications, University of Electronic Science and Technology of
China, Chengdu 611731, China (e-mail: lzhou@uestc.edu.cn)}
}


\maketitle

\begin{abstract}

Consider the computations at a node in a message passing algorithm.
Assume that the node has incoming and outgoing messages $\mb{x}  = (x_1, x_2, \ldots, x_n)$ and $\mb{y} = (y_1, y_2, \ldots, y_n)$, respectively.
In this paper, we investigate a class of structures that can be adopted by the node for computing $\mb{y}$ from $\mb{x}$, where each $y_j, j = 1, 2, \ldots, n$ is computed via a binary tree with leaves $\mb{x}$ excluding $x_j$.
We make three main contributions regarding this class of structures.
First, we prove that the minimum complexity of such a structure is $3n - 6$, and if a structure has  such complexity, its minimum latency is $\delta + \lceil \log(n-2^{\delta}) \rceil$ with $\delta = \lfloor \log(n/2) \rfloor$, where the logarithm always takes base two.
Second, we prove that the minimum latency of such a structure is $\lceil \log(n-1) \rceil$, and if a structure has such latency, its minimum complexity is $n \log(n-1)$ when $n-1$ is a power of two.
Third, given $(n, \tau)$ with $\tau \geq \lceil \log(n-1) \rceil$, we propose a construction for a structure which we conjecture to have the minimum complexity among structures with latencies at most $\tau$.
Our construction method runs in $O(n^3 \log^2(n))$ time, and the obtained structure has complexity at most (generally much smaller than) $n \lceil \log(n) \rceil - 2$.
\end{abstract}


\IEEEpeerreviewmaketitle


\section{Introduction}\label{section: introduction}

Message passing algorithms are widely applied for the decoding of error correction codes such as the low-density parity-check (LDPC) codes \cite{	Gallager62, Richardson01capacity, Chen05, he2019onmutual, he2019mutual, he2019onfinite}.
The algorithms can be considered as working on a graph, in which messages are passing along edges, and each node receives incoming messages from its connecting edges and then computes outgoing messages that will be passed back along the connecting edges.
More specifically, consider a node, such as a check/variable node of LDPC codes, which has $n$ connecting edges.
(We assume $n \geq 3$ throughout this paper and specify cases for $n < 3$ separately.)
The incoming messages are denoted by $\mb{x}  = (x_1, x_2, \ldots, x_n)$, where for $j \in [n] = \{1, 2, \ldots, n\}$, $x_j$ comes from the $j$-th connecting edge.
This node then computes $n$ outgoing messages, denoted by $\mb{y} = (y_1, y_2, \ldots, y_n)$, where for $j \in [n]$, $y_j$ will be passed back along the $j$-th connecting edge.
The corresponding node computations are to compute each $y_j, j \in [n]$ from $\mb{x}$ excluding $x_j$.
We remark that the messages need not to be real numbers.

\begin{figure}[t]
\centering
\includegraphics[scale = 0.5]{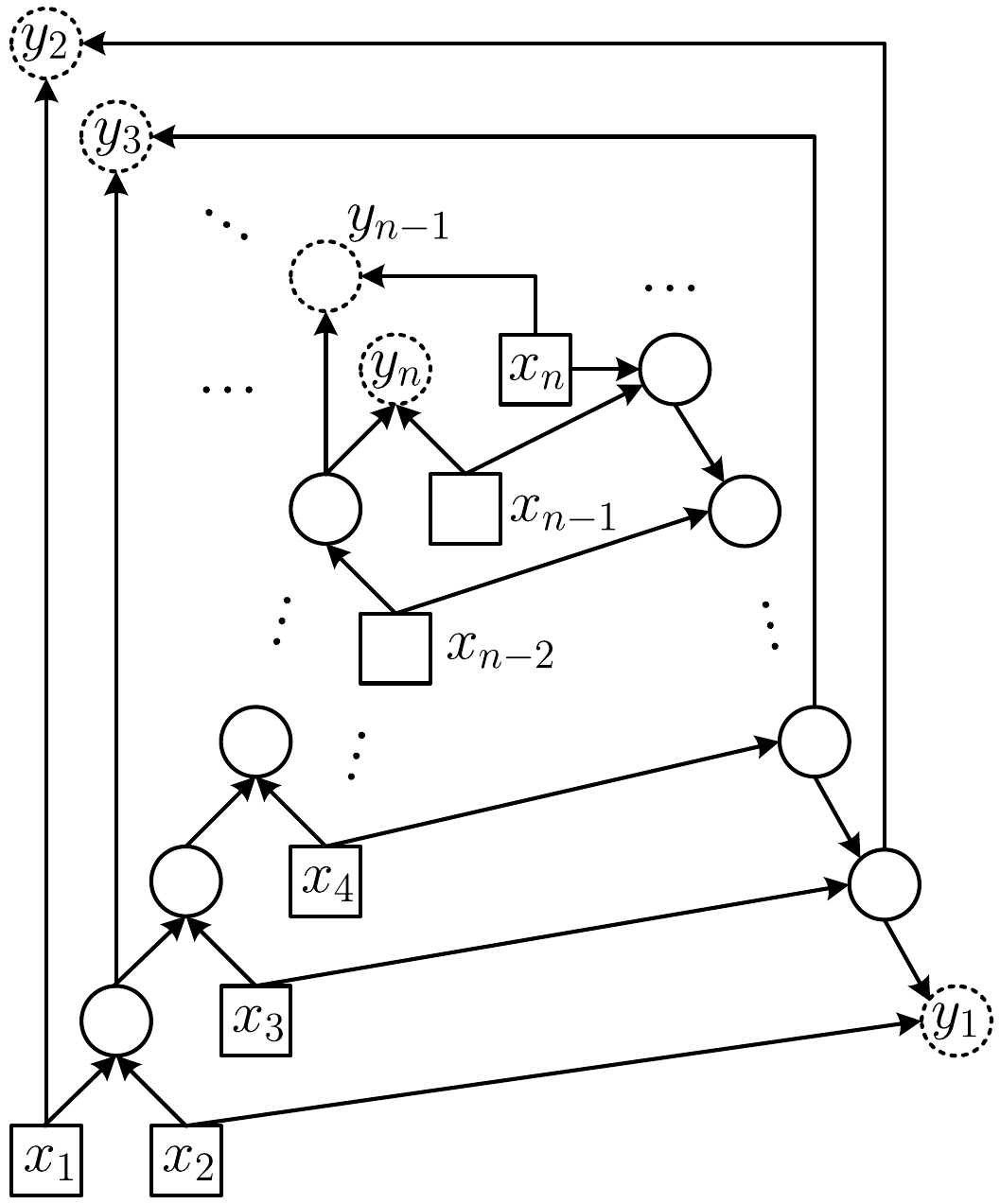}
\caption{A structure for realizing the forward-backward computation of $\mb{y}$ \cite{hu2001efficient}, where squares, circles, and dotted circles represent input, internal, and output nodes, respectively.}
\label{fig: for-back}
\end{figure}

\begin{figure}[t]
\centering
\includegraphics[scale = 0.5]{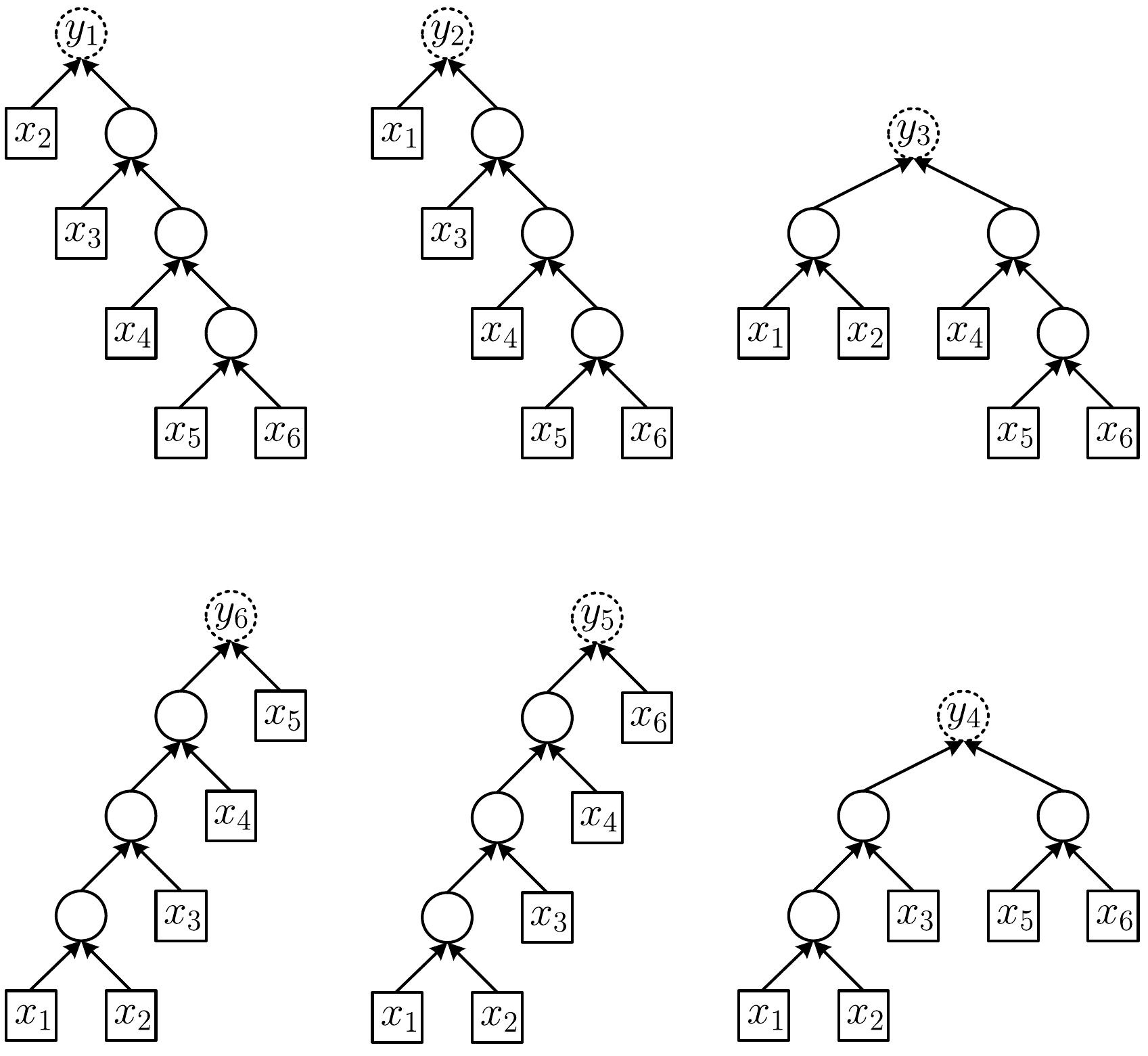}
\caption{Examples of directed binary trees (DBTs) used for computing $y_j, \forall j \in [n] = [6]$, where squares, circles, and dotted circles represent leaves, internal nodes, and roots, respectively.
}
\label{fig: DBT}
\end{figure}

In this paper, we consider a class of structures, in which each $y_j, j \in [n]$ is computed by using a binary tree with leaves $\mb{x}$ excluding $x_j$.
For example, assume
\begin{equation}\label{eqn: yj = min}
y_j = \min_{i \in [n]\setminus\{j\}} x_i, \forall j \in [n],
\end{equation}
which is used in the computation at the check node in the min-sum decoding of LDPC codes \cite{	Chen05} (messages considered here are real numbers).
Fig. \ref{fig: for-back} shows a classical structure \cite{	hu2001efficient} for the computation of \eqref{eqn: yj = min}.
This structure realizes the computation of a given $y_j$ based on a binary tree whose leaves correspond to $\{x_i: i \in [n]\setminus\{j\} \}$ and whose internal nodes correspond to the two-input $\min$ operations.
Taking $n = 6$ as an example, the six binary trees resulted from Fig. \ref{fig: for-back} are shown in Fig \ref{fig: DBT}.

The structure in Fig. \ref{fig: for-back} actually carries out the forward-backward computation \cite{hu2001efficient}.
Taking the computation of \eqref{eqn: yj = min} as an example, the forward and backward computations are given by
\[
    f_j = \min_{i = 1, \ldots, j} x_i =
    \begin{cases}
        x_1, & j = 1,\\
        \min\{f_{j-1}, x_j\}, & 1 < j < n,
    \end{cases}
    \quad \text{and}
\]
\[
    b_j = \min_{i = j, \ldots, n} x_i =
    \begin{cases}
        x_n, & j = n,\\
        \min\{b_{j+1}, x_j\}, & 1 < j < n,
    \end{cases}
\]
respectively.
Then, we have
\[
    y_j = \min_{i \in [n] \setminus \{j\}} x_i =
    \begin{cases}
        b_{j+1}, & j = 1,\\
        \min\{f_{j-1}, b_{j+1}\}, & 1 < j < n,\\
        f_{j-1}, & j = n.
    \end{cases}
\]
The complexity of this structure is defined as the number of internal nodes ($\min$ operations) which is given by $3n - 6$: each of $\{f_2, \ldots, f_{n-1}, b_2, \ldots, b_{n-1}, y_2, \ldots, y_{n-1}, \}$ takes one $\min$ operation.
The latency of the structure is defined as the longest distance between any pair of $(x_i, y_j), i \neq j$, which is given by $n - 2$ (e.g., from $x_1$ to $y_n$).

It is natural to ask what are the minimum complexity and minimum latency of such a class of structures?
Accordingly, this paper derives the following results.
\begin{itemize}
\item   We prove that the minimum complexity of such a structure is $3n - 6$.
    If a structure has such complexity (i.e., complexity-optimal), its minimum latency is $\delta + \lceil \log(n-2^{\delta}) \rceil$ with $\delta = \lfloor \log(n/2) \rfloor$, where the logarithm always takes base two in this paper.
    We also propose a simple construction for complexity-optimal structures which have such  latency.
\item   We prove that the minimum latency of such a structure is $\lceil \log(n-1) \rceil$.
    If a structure has such latency (i.e., latency-optimal), its minimum complexity is $n \log(n-1)$ for $n = 2^k + 1$ with $k > 0$, and we propose a simple construction for this case.
\item   Given $(n, \tau)$ with $\tau \geq \lceil \log(n-1) \rceil$, we propose a construction for a structure $S_{n, \tau}$  which we conjecture to have the minimum complexity among structures with latencies at most $\tau$.
    Our construction method runs in $O(n^3 \log^2(n))$ time, and the obtained $S_{n, \tau}$ has complexity at most (generally much smaller than) $n \lceil \log(n) \rceil - 2$.
\end{itemize}

The complexities of $S_{n, \tau}$, denoted by $\phi (n, \tau)$, are derived in Section \ref{section: Tradeoff}, and  some typical values of $\phi(n, \tau)$ are presented in Table \ref{table: phi} of Section \ref{section: Tradeoff}.
It is worth mentioning that structures that are both complexity-optimal and latency-optimal only exist for $n = 2, 3, 4, 6$.
Assume that the min-sum algorithm \cite{	Chen05} (or its variants) is applied to decode the 802.11n LDPC code \cite{IEEESTD802_11n} which has check node degrees of 7 and 8.
For each degree-7 check node ($n = 7$), using the structure of Fig. \ref{fig: for-back} to implement \eqref{eqn: yj = min} leads to complexity 15 and latency 5.
On the contrary, Table \ref{table: phi} shows that there exist a structure of complexity 15 and latency 4, and also a structure of complexity 18 and latency 3.
Moreover, for each degree-8 check node ($n = 8$), using the structure of Fig. \ref{fig: for-back} for implementing \eqref{eqn: yj = min} results in complexity 18 and latency 6.
Table \ref{table: phi}, however, shows that there exist a structure of complexity 18 and latency 4, and a structure of complexity 22 and latency 3.

We remark that there exist some other structures \cite{wey2008algorithms, lee2015low} which are specially designed for the computation of \eqref{eqn: yj = min}.
They do not belong to the class of structures considered in this paper.
To make a fair comparison in terms of complexity and latency, more factors need to be taken into consideration: the comparators with different bit widths, the multiplexers, the latency of comparators and multiplexers, and so on, which are out of the scope of the current paper.
We thus only make two more remarks.
First, the structures proposed in \cite{wey2008algorithms, lee2015low} can  never achieve the minimum latency $\lceil \log(n-1) \rceil$.
Second, they are only suitable for the $\min$ operations, while the class of structures considered in this paper is always applicable for the node computation no matter what binary operations are involved.
For example, the considered class of structures is perfectly suitable for the mutual information-maximizing lookup table (MIM-LUT) decoding \cite{Kurkoski08, Romero15decoding, Romero16, meidlinger2020design, Meidlinger15, Lewandowsky16, balatsoukas2015fully, ghanaatian2018a588, Lewandowsky18, lewandowsky2018design, lewandowsky2019design}, which recently attracts much attention as it takes two-input table lookup operations to eliminate arithmetic operations.

The remainder of this paper is organized as follows.
Section \ref{section: preliminary} introduces preliminaries regarding graphs and trees.
Section \ref{section: Structures} defines the structures considered in this paper for node computation.
Sections \ref{section: Complexity-Optimal} and \ref{section: Latency-Optimal} investigate complexity-optimal and latency-optimal structures, respectively.
Section \ref{section: Tradeoff} considers the construction of the aforementioned structure $S_{n, \tau}$.
Finally, Section \ref{section: Conclusion} concludes this paper.

\section{Preliminaries}\label{section: preliminary}

In this section, we introduce preliminaries regarding graphs and trees, mainly based on their definitions in \cite[Appendix B]{cormen2009introduction}.

A directed (resp. undirected) graph $G$ is a pair $(G_v, G_e)$, where $G_v$ and $G_e$ are the node/vertex set and edge set, respectively, and any element in $G_e$ is called a directed (resp. undirected) edge which is denoted by an ordered (resp. unordered) pair $(a, b) \in G_v \times G_v$.
The term ``ordered" (resp. ``unordered") implies that $(a, b) \neq (b, a)$ (resp. $(a, b) = (b, a)$). (Note that self-loops are forbidden in this paper, i.e., we have $a \neq b, \forall (a, b) \in G_e$.)
When drawing a graph, we use arrows and lines to represent directed and undirected edges, respectively.
For convenience, we also consider that $G = G_v \cup G_e$, and accordingly, we also write $a \in G_v$ as $a \in G$ and $(a, b) \in G_e$ as $(a, b) \in G$.
A graph $G'$ is called a subgraph of $G$ if $G' \subseteq G$.

In a directed graph $G$, we say that $(a, b) \in G$ leaves $a$ and enters $b$; accordingly, $(a, b)$ is a leaving/outgoing edge of $a$ and an entering/incoming edge of $b$.
Instead, in an undirected graph $G$, we simply say that $(a, b) \in G$ connects $a$ and $b$; accordingly $(a, b)$ is an edge of $a$ and $b$.
We use the subtraction/addition (i.e., $-/+$) to describe the operation of removing/adding a node $a$ or an edge  $(a, b)$ from/into a graph $G = (G_v, G_e)$, where $(a, b)$ is a directed edge if and only if (iff) $G$ is a directed graph.
Specifically, $G - a = (G_v\setminus\{a\}, G_e\setminus\{(a_1, a_2) \in G_e: a_1 = a ~\text{or}~ a_2 = a\})$, $G + a = (G_v \cup \{a\}, G_e)$, $G - (a, b) = (G_v, G_e \setminus \{(a, b)\})$, and $G + (a, b) = (G_v\cup\{a, b\}, G_e \cup \{(a, b)\})$.

A path $P$ of length $k$ from $a \in G$ to $b \in G$ is a node sequence $P = (v_0, v_1, \ldots, v_k)$ such that $v_0 = a$, $v_k = b$, and $(v_{i-1}, v_i) \in G, \forall i \in [k]$.
The distance from $a$ to $b$ is the length of the shortest path from $a$ to $b$ (the distance is defined as $\infty$ if there is no such a path).
$P$ is a simple path if $v_i \neq v_j, \forall~ 0 \leq i < j \leq k$.
Moreover, $P$ forms a (simple) cycle if $k \geq 2$, $v_0 = v_k$, $(v_0, v_1) \neq (v_1, v_2)$, and $(v_1, v_2, \ldots, v_k)$ is a simple path.
A graph with no cycle is acyclic.
We refer to a directed acyclic graph by a DAG.
If there is a path $P$ from $a$ to $b$, we say that $b$ is reachable from $a$ (via $P$), denoted by $a \rightsquigarrow b$.
For any directed graph $G$ and $a \in G$, we say that $E(a,G) = (\{b \in G: b \rightsquigarrow a\}, \{(b', b'') \in G: b'' \rightsquigarrow a\} )$ is the subgraph entering $a$ in $G$, and $L(a,G) = (\{b \in G: a \rightsquigarrow b\}, \{(b', b'') \in G: a \rightsquigarrow b'\} )$ is the subgraph leaving $a$ in $G$.
An undirected graph is connected if every node is reachable from all other nodes.

A tree is a connected, acyclic, undirected graph.
For any tree $T$ and any node $a \in T$, $a$ is called an internal node  (resp. external node or leaf) if $a$ has more than one (resp. only one) edge.
There is a unique simple path between any pair of nodes in $T$.
The diameter of $T$, denoted by $d(T)$, is the length of the longest simple path in $T$.

A rooted tree is a tree in which there is a unique node called the root of the tree.
Consider a rooted tree $T$, and denote its root by $r(T)$.
The distance between any node $a \in T$ and $r(T)$ is called the depth of $a$ in $T$.
A level of $T$ consists of all  nodes at the same depth.
The height of $T$ is equal to the largest depth of any node in $T$.
For any edge $(a, b) \in T$, assuming that $a$ has a larger depth (which is equal to one plus the depth of $b$), then, $b$ is called the parent of $a$, and $a$ is called a child of $b$.
The directed version of $T$, say $T'$, is to change each undirected edge, say $(a, b) \in T$ with $a$ being a child of $b$, into the directed edge $(a, b) \in T'$.
$T'$ is called a directed rooted tree (DRT), and we say that $T$ is the undirected version of $T'$.
For any $a \in T'$, $E(a, T')$ is the subtree of $T'$ rooted at $a$; accordingly, the undirected version of $E(a, T')$ is the subtree of $T$ rooted at $a$.

A (full) binary tree $T$ is a rooted tree in which each node has either zero or two children (left child and right child).
Assume that the height of an arbitrary binary tree $T$ is $h_T$.
$T$ is called a complete binary tree iff for $0 \leq i < h_T$, the $i$-th level of $T$ contains $2^i$ nodes, and nodes in the $h_T$-th level of $T$ are as far left as possible.
Moreover, $T$ is called a perfect binary tree iff for $0 \leq i \leq h_T$, the $i$-th level of $T$ contains $2^i$ nodes.
The subtree rooted at the left (resp. right) child of $r(T)$ is called the left (resp. right) subtree of $T$.
Similar to DRTs, we have directed binary trees (DBTs).
Meanwhile, we refer to the directed version of a complete (resp. perfect) binary tree as a complete (resp. perfect) DBT.

For any graph $G = (G_v, G_e)$, $G$ is labelled iff every node in $G$ is given a unique label, such as $1, 2, \ldots, |G_v|$ (as a result, each edge is also given a unique label).
Otherwise, $G$ is partially unlabelled (even if no node is labelled).
Two labelled graphs $G$ and $G'$ are the same, i.e., $G = G'$, iff $G$ and $G'$ have same labelled nodes and edges (and root for rooted trees).
Two partially unlabelled graphs $G$ and $G'$ are the same iff there exists a way to label all unlabelled nodes in $G$ and $G'$ such that $G$ and $G'$ become labelled and the same.

\section{Structures for Node Computation}\label{section: Structures}

Recall that $\mb{x}  = (x_1, x_2, \ldots, x_n)$ and $\mb{y} = (y_1, y_2, \ldots, y_n)$ denote the incoming and outgoing messages, respectively.
In this paper, we consider the case where for $j \in [n]$, a DBT $T_j$ is used to describe the computation of $y_j$ from $\mb{x}$ excluding $x_j$.
More specifically, in $T_j$, leaves correspond to incoming messages $\mb{x}$ excluding $x_j$, internal nodes correspond to binary operations, and the root $r(T_j)$ corresponds to $y_j$.
Some examples of such DBTs for $n = 6$ are shown in Fig. \ref{fig: DBT}.

Define an input node set $X = \{x_j: j \in [n]\}$ and an output node set $Y = \{y_j: j \in [n]\}$, where node $x_j$ (resp. $y_j$) is called the $j$-th input (resp. output) node which corresponds to the $j$-th incoming message $x_j$ (resp. outgoing message $y_j$).
In this paper, we remark that for any graph $G$ and any node $a \in G$, $a$ is labelled in $G$ iff $a$ is an input node from $X$ or an output node from $Y$.
As a result, $G$ is generally partially unlabelled.
We consider to use a structure, defined below, to describe a computation process.

\begin{definition}\label{definition: structure}
A structure $S$ considered in this paper is a DAG fulfilling the following three properties.
\begin{itemize}
\item   For any $a \in S$, we have $a \in X$ iff $a$ has no incoming edge in $S$.
\item   For any $a \in S$, $E(a, S)$ is a DBT.
\item   For two different nodes $a, b \in S$, $E(a, S) \neq E(b, S)$ (the inequality corresponds to comparison between two partially unlabelled graphs).
\end{itemize}
\end{definition}

For any $a \in S \cap X$, we also call $a$ an input node of $S$, and we say that $S$ has input size $|S \cap X|$ (the number of input nodes in $S$).
Any other node in $S$ is called a computation node, and it must have exactly two incoming edges in $S$.
In particular, any computation node with no outgoing edge is also called an output node (may not belong to $Y$).
For any $a \in S$, we call $E(a, S)$ the subtree of $S$ rooted at $a$.
The third property in Definition \ref{definition: structure} indicates that $S$ does not have the same subtrees.
For convenience, let $E(S) = \{E(a, S): a \in S\}$ be the set of all subtrees of $S$.
For any two structures $S$ and $S'$, denote the union of $S$ and $S'$ by $S \vee S'$, where only one copy of the same subtrees is kept such that $S \vee S'$ is still a structure.
We have $E(S \vee S') = E(S) \cup E(S')$.
For example, the six DBTs (structures) in Fig. \ref{fig: DBT} can be united (under $\vee$) into the structure shown in Fig. \ref{fig: for-back} with $n = 6$.

\begin{definition}\label{definition: structure y}
A structure $S$ used for computing $\mb{y}$ is a structure (see Definition \ref{definition: structure}) additionally fulfilling the following property.
\begin{itemize}
\item   $S$ contains $n$ output nodes, which are exactly $Y = \{y_1, y_2, \ldots, y_n\}$, where for $j \in [n]$, $E(y_j, S)$ is a DBT with leaves $X \setminus \{{x_j}\}$.
\end{itemize}
\end{definition}

\begin{figure}[t]
\centering
\includegraphics[scale = 0.5]{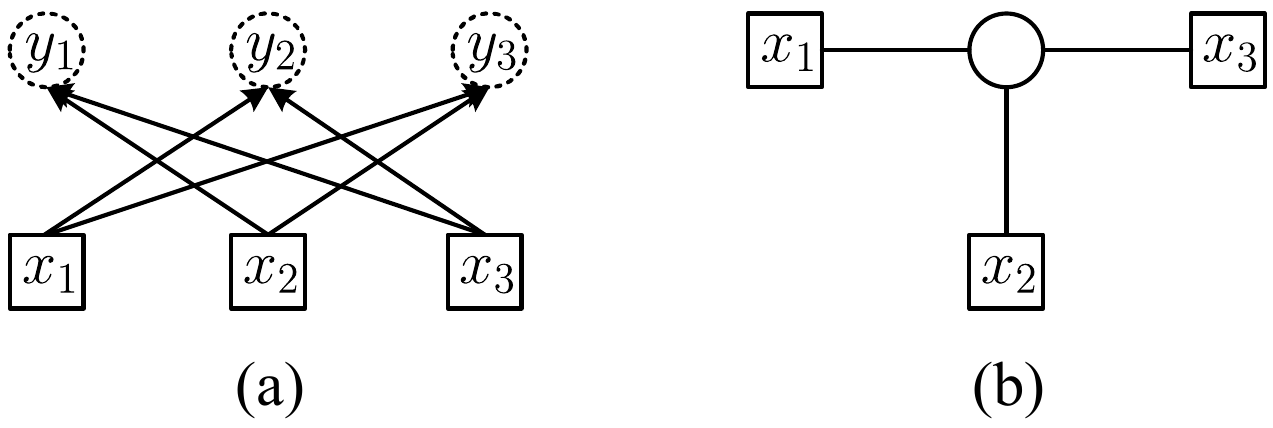}
\caption{(a) The only structure $S \in \mc{S}_3$ and (b) only T-tree $T \in \mc{T}_3$, and we have $S = h(T)$.
}
\label{fig: S3}
\end{figure}

We remark that any structure defined by Definition \ref{definition: structure y} can be used for computing $\mb{y}$, but that defined by Definition \ref{definition: structure} may not.
From Definition \ref{definition: structure y}, we have $S = \vee_{j \in [n]} E(y_j, S)$, and $S$ has input size $n$.
Let $\mc{S}_n$ be the set of all structures used for computing $\mb{y}$ (and with input size $n$).
Fig. \ref{fig: for-back} shows an instance in $\mc{S}_n$.
Meanwhile, the only structure in $\mc{S}_3$ is shown by Fig. \ref{fig: S3}(a).
For any $S \in \mc{S}_n$, it is easy to see that for any $j \in [n]$, each output node in $Y \setminus \{y_j\}$ is reachable from input node $x_j$ via a unique path in $S$, but $y_j$ is not reachable from $x_j$.
Moreover, after removing any nodes and/or edges from $S$, we can no longer have $S \in \mc{S}_n$. 

\begin{definition}\label{definition: complexity latency}
For any structure $S$, the complexity of $S$, denoted by $c(S)$, is equal to the number of computation nodes in $S$.
The latency of $S$, denoted by $l(S)$, is equal to the length of the longest simple path in $S$.
\end{definition}

As an example, the complexity and latency of the structure in Fig. \ref{fig: for-back} are $3n - 6$ and $n - 2$, respectively.
It is reasonable to use complexity and latency as two key criteria for evaluating the performance of a structure.
In this paper, one of our main purpose is to discover complexity-optimal and/or latency-optimal structures in $\mc{S}_n$, as defined below.

\begin{definition}\label{definition: optimal}
Let $c_n^{\min} = \min_{S \in \mc{S}_n} c(S)$ and $l_n^{\min} = \min_{S \in \mc{S}_n} l(S)$.
Moreover, let $\mc{S}_n^{\text{co}} = \{S \in \mc{S}_n: c(S) = c_n^{\min}\}$ and $\mc{S}_n^{\text{lo}} = \{S \in \mc{S}_n: c(S) = c_n^{\min}\}$.
For any structure $S \in \mc{S}_n$, $S$ is complexity-optimal (resp. latency-optimal) iff $S \in \mc{S}_n^{\text{co}}$ (resp. $S \in \mc{S}_n^{\text{lo}}$).
\end{definition}

\section{Complexity-Optimal Structures}\label{section: Complexity-Optimal}

In this section, we first investigate the properties of  complexity-optimal structures, including deriving the value of $c_n^{\min}$.
Then, we propose to use a class of trees, called T-trees, to equivalently describe complexity-optimal structures.
T-trees make it easy to find the minimum latency of complexity-optimal structures, and also lead to a simple construction for complexity-optimal structures.

\subsection{Properties of Complexity-Optimal Structures}

\begin{figure}[t]
\centering
\includegraphics[scale = 0.5]{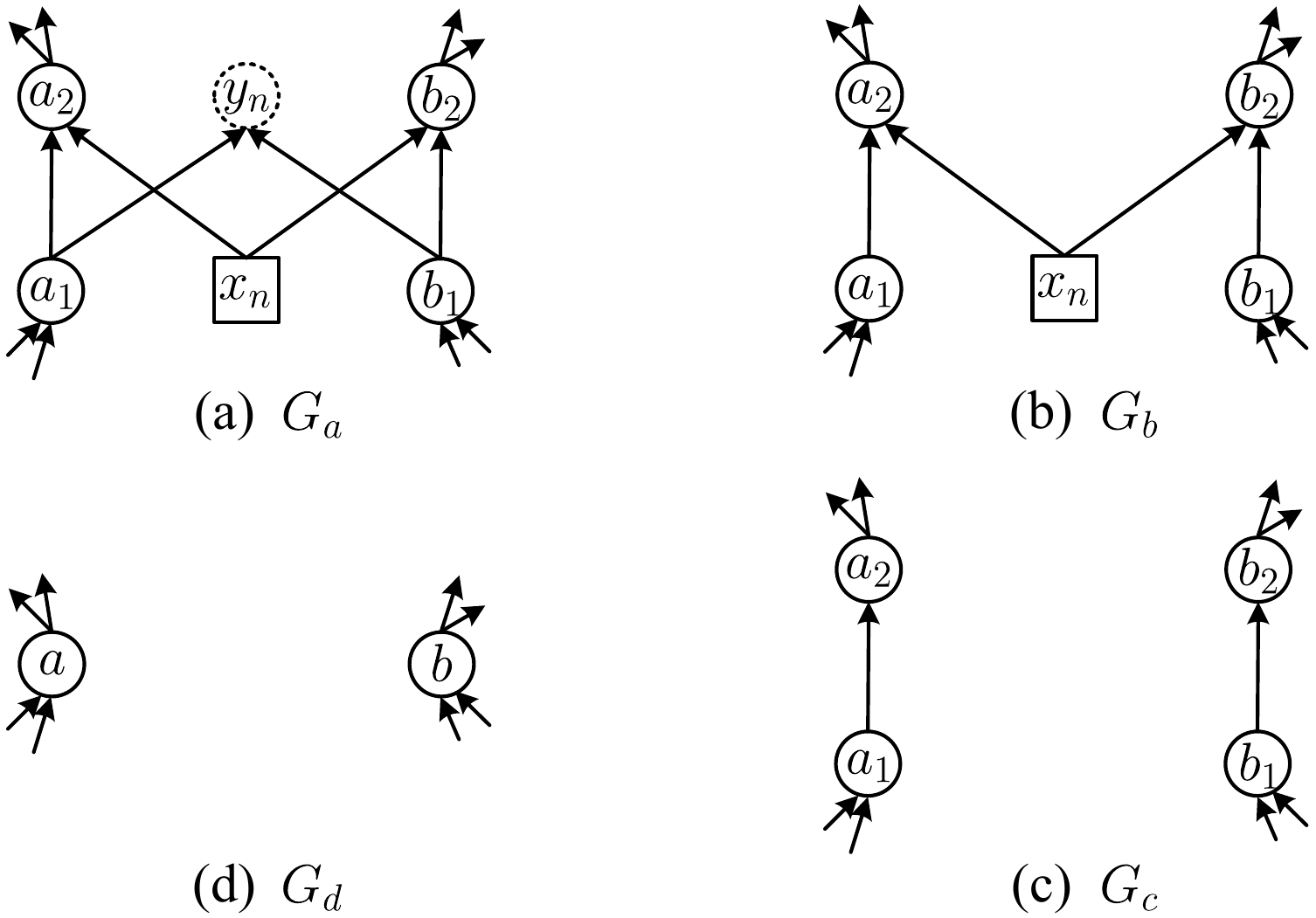}
\caption{An example for illustrating functions $f$ and $g$. (For simplicity, only the subgraphs of interest, but not the whole graphs, are drawn.)
$G_d = f(G_a)$: (I) $G_b = G_a - y_n$, (II) $G_c = G_b - x_n$, and (III) $G_d$ is the resulting graph after converting $(a_1, a_2)$ and $(b_1, b_2)$ in $G_c$ into $a$ and $b$, respectively. (Steps (I)--(III) correspond to steps (A1)--(A3).)
On the other hand, $G_a = g(a, b, G_d)$: (i) $G_c$ is the resulting graph after converting $a$ and $b$ in $G_d$ into $(a_1, a_2)$ and $(b_1, b_2)$, respectively, (ii) $G_b = G_c + (x_n, a_2) + (x_n, b_2)$, and (iii) $G_a = G_b + (a_1, y_n) + (b_1, y_n)$. (Steps (i)--(iii) correspond to steps (B1)--(B3).)
}
\label{fig: xn_Sn}
\end{figure}

For any directed graph $G$ and any node $a \in G$, converting $a$ into a directed edge $(a_1, a_2)$ is to split $a$ into two new nodes $a_1$ and $a_2$ in $G$ such that they keep only the incoming and outgoing edges of $a$, respectively, and the directed edge $(a_1, a_2)$ is also added into $G$ ($a$ no longer exists in $G$).
An example is shown in Fig. \ref{fig: xn_Sn}, where $G_c$ is the resulting graph after converting $a$ and $b$ in $G_d$ into $(a_1, a_2)$ and $(b_1, b_2)$, respectively.
Conversely, for $(a_1, a_2) \in G$, converting $(a_1, a_2)$  into node $a$ is to merge $a_1$ and $a_2$ into the new node $a$ in $G$ such that $a$ keeps all edges of $a_1$ and $a_2$ except for the edge $(a_1, a_2)$ (nodes $a_1$ and $a_2$ no longer exist in $G$).
An example is also shown in Fig. \ref{fig: xn_Sn}, where $G_d$ is the resulting graph after converting $(a_1, a_2)$ and $(b_1, b_2)$ in $G_c$ into $a$ and $b$, respectively.

For $n \geq 4$, define a function $f: \mc{S}_n \to \mc{S}_{n-1}$ which works with the following three steps for any $S \in \mc{S}_n$. (An example is shown in Fig. \ref{fig: xn_Sn} to illustrate how $f$ works.)
\begin{enumerate}[(\text{A}1)]
\item    For any output node $a \in S$ (i.e., $a$ has no outgoing edge) such that $a \notin Y\setminus\{y_n\}$, remove $a$ from $S$ in a recursive manner. Denote the resulting graph by $S'$.
\item   Let $S'' = S' - x_n$.
\item   For any $(a, b) \in S''$ such that $(a, b)$ is the only incoming edge of $b$ in $S''$, convert $(a, b)$ into a new node. (Actually, $b$ is a computation node in $S$ with incoming edges $(a, b)$ and $(x_n, b)$.) Denote the resulting graph by $f(S)$.
\end{enumerate}

\begin{lemma}\label{lemma: c(S) >= c(f(S)) + 3}
For $n \geq 4$ and any $S \in \mc{S}_n$, we have $f(S) \in \mc{S}_{n-1}$ and $c(S) \geq c(f(S)) + 3$.
\end{lemma}

\begin{IEEEproof}
Assume $n \geq 4$ and $S \in \mc{S}_n$.
We can easily verify that $f(S) \in \mc{S}_{n-1}$.
In step (A1) of $f(S)$, at least the computation node $y_n$ is removed from $S$.
The number of additional computation nodes removed from $S$ in steps (A2) and (A3) is equal to the number of edges of $x_n$ in $S$.
Therefore, to prove $c(S) \geq c(f(S)) + 3$, we only need to prove that $x_n$ has at least two outgoing edges in $S$.

Since each output node in $Y\setminus \{y_n\}$ is reachable from $x_n$ in $S$, $x_n$ must have at least one outgoing edge, say $(x_n, a) \in S$.
Note that $a$ is a computation node in $S$, indicating that $a$ is reachable from at least an input node $x_i, i \neq n$ in $S$.
Since $y_i$ is not reachable from $x_i$ in $S$, then $y_i$ must not be reachable from $a$.
Therefore, $x_n$ must have another outgoing edge such that $y_i$ can be reachable from $x_n$ in $S$.
This completes the proof.
\end{IEEEproof}

\begin{theorem}\label{theorem: c_n^min = 3n - 6}
We have $c_n^{\min} = \min_{S \in \mc{S}_n} c(S) = 3n - 6$.
\end{theorem}

\begin{IEEEproof}
We have $c_n^{\min} = 3n - 6$ for $n = 3$, since $\mc{S}_3$ contains only one structure, as shown in Fig. \ref{fig: S3}(a).
Then, according to Lemma \ref{lemma: c(S) >= c(f(S)) + 3}, we have $c_n^{\min} \geq 3n - 6$ for $n \geq 4$.
Further noting that the structure in Fig. \ref{fig: for-back} has complexity $3n - 6$, the theorem is proved.
\end{IEEEproof}

According to the discussions on Lemma \ref{lemma: c(S) >= c(f(S)) + 3} and Theorem \ref{theorem: c_n^min = 3n - 6}, we know that for $n \geq 4$ and any $S \in \mc{S}_n^{\text{co}}$, we have $c(S) = c(f(S)) + 3$.
More specifically, only one computation node, i.e., $y_n$, is removed in step (A1) of $f(S)$, $x_n$ has exactly two outgoing edges in $S$, and we have $f(S) \in \mc{S}_{n-1}^{\text{co}}$.
This motivates us to construct another function, which works like the inverse process of $f$, to convert a structure in $\mc{S}_{n-1}^{\text{co}}$ to a structure in $\mc{S}_{n}^{\text{co}}$.

For any $a, b \in S \in \mc{S}_{n}$, the unordered pair $\langle a, b \rangle $ is called a complement pair of $S$ iff $x_j \in E(a, S) \iff x_j \notin E(b, S), \forall j \in [n]$.
For example, $\langle x_3, y_3 \rangle  = \langle y_3, x_3\rangle $ is a complement pair of $S$ in Fig. \ref{fig: S3}(a).
Let $P(S)$ denote the set of all complement pairs of $S$.
For $n \geq 4$, $S \in \mc{S}_{n-1}^{\text{co}}$, and $\langle a, b\rangle \in P(S)$, define $g(a, b, S)$ as the graph obtained by the three steps described as follows. (An example is shown in Fig. \ref{fig: xn_Sn} to illustrate how $g$ works.)
\begin{enumerate}[(B1)]
\item   Convert $a$ and $b$ into directed edges $(a_1, a_2)$ and $(b_1, b_2)$, respectively. Denote the resulting graph by $S'$.
\item   Let $S'' = S' + (x_n, a_2) + (x_n, b_2)$.
\item   Let $g(a, b, S) = S'' + (a_1, y_n) + (b_1, y_n)$.
\end{enumerate}

\begin{theorem}\label{theorem: Sn property}
Structures in $\mc{S}_n^{\text{co}}$ fulfill the following properties.
\begin{enumerate}[(C1)]
\item   For any $S \in \mc{S}_{n}^{\text{co}}$, any non-output node in $S$ has exactly two outgoing edges.
\item   For any $S \in \mc{S}_{n}^{\text{co}}$, $a \in S$, and $j \in [n]$, we have $x_j \in E(a, S) \iff y_j \notin L(a, S)$.
\item   For any $S \in \mc{S}_{n}^{\text{co}}$ and $a \in S$, there exists a unique $b \in S$ such that $\langle a, b \rangle \in P(S)$.
\item   For $n \geq 4$, $\mc{S}_{n}^{\text{co}} = \{g(a, b, S): S \in \mc{S}_{n-1}^{\text{co}}, \langle a, b \rangle \in P(S)\}$.
\item   $|\mc{S}_{n}^{\text{co}}| = (2n - 5)!! = (2n-5)\times(2n-7)\times\cdots\times1$.
\end{enumerate}
\end{theorem}

\begin{IEEEproof}
See Appendix \ref{appendix: Sn property}.
\end{IEEEproof}

\subsection{T-Trees}

\begin{definition}\label{definition: T-tree}
A T-tree $T$ used for computing $\mb{y}$ is a (undirected) tree fulfilling the following two properties.
\begin{itemize}
\item   $T$ has $n$ leaves, which are exactly $X$.
\item   Each internal node in $T$ has exactly three edges.
\end{itemize}
\end{definition}

The letter `T' in ``T-tree'' actually comes from the second property above (`T' is short for ``Triplet", and it also looks like an internal node with three edges).
Denote $\mc{T}_n$ as the set of all T-trees.
In particular, the only T-tree in $\mc{T}_3$ is shown in Fig. \ref{fig: S3}(b).
For any $T \in \mc{T}_n$, $T$ has $n - 2$ internal nodes and $2n - 3$ edges.
For any $(a, b) \in T$, let $D(a, b, T) = E(a, T_b)$, where $T_b$ is the directed version of the tree resulted by making $T$ as a rooted tree with root $b$.
Obviously, $D(a, b, T)$ is a DBT with root $a$.
Let $D(T) = \{D(a, b, T): (a, b) \in T\}$.
We have $|D(T)| = 4n - 6$, since $D(a_1, b_1, T) \neq D(a_2, b_2, T)$ for any $(a_1, b_1), (a_2, b_2) \in T$ with $a_1 \neq a_2$ or $b_1 \neq b_2$.

\begin{theorem}\label{theorem: h bijective}
For any $T \in \mc{T}_n$, let
\[
    h(T) = \vee_{j \in [n], (a, x_j) \in T} D(a, x_j, T).
\]
Then $h$ is a bijection from $\mc{T}_n$ to $\mc{S}_n^{\text{co}}$.
\end{theorem}

\begin{IEEEproof}
See Appendix \ref{appendix: h bijective}.
\end{IEEEproof}

\begin{figure}[t]
\centering
\includegraphics[scale = 0.5]{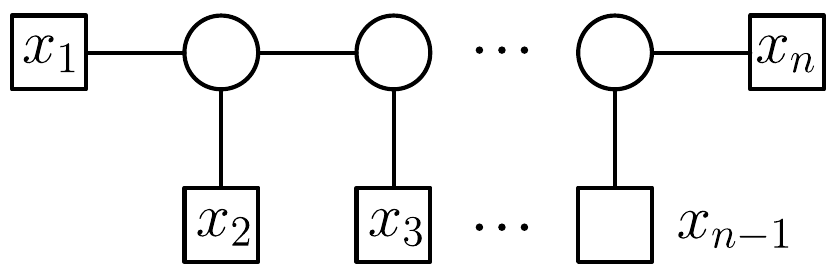}
\caption{The T-tree $T \in \mc{T}_n$ with $h(T) \in \mc{S}_n^{\text{co}}$ given by Fig. \ref{fig: for-back}.
}
\label{fig: for-back-CT}
\end{figure}

According to Theorem \ref{theorem: h bijective}, it suffices to investigate $\mc{T}_n$ when $\mc{S}_n^{\text{co}}$ is of interest.
In particular, for any $T \in \mc{T}_n$, $T$ is a  much simpler graph than $h(T) \in \mc{S}_n^{\text{co}}$, with respect to that i) $T$ is a simple tree as described in Definition \ref{definition: T-tree} and ii) $T$ contains $2n-2$ nodes and $2n-3$ edges while $h(T)$ contains $4n-6$ nodes and $6n - 12$ edges.
As an example, let $S$ denote the structure in Fig. \ref{fig: for-back}, and we have $S \in \mc{S}_n^{\text{co}}$.
The T-tree $T$ with $h(T) = S$ is shown in Fig. \ref{fig: for-back-CT}.
A simpler example for $h$ is shown in Fig. \ref{fig: S3}.

\begin{lemma}\label{lemma: l(h(T)) = d(T)-1}
For any $T \in \mc{T}_n$, we have $l(h(T)) = d(T) - 1$, where $d(T)$ is the diameter of $T$.
\end{lemma}

\begin{IEEEproof}
Note that $d(T)$ must be equal to the distance between a certain pair of leaves in $T$.
Without loss of generality, assume that $d(T)$ is equal to the distance between $x_i$ and $x_j$ in $T$.
As a result, $D(a, x_j, T)$ with $(a, x_j) \in T$ has the largest height among $D(T)$.
Therefore, $l(h(T))$ is equal to the height of $D(a, x_j, T)$, i.e., $l(h(T)) = d(T) - 1$.
\end{IEEEproof}

\begin{lemma}\label{lemma: d_n^min}
Let $\delta = \lfloor \log(n/2) \rfloor$.
We have
\[
    d_n^{\min} = \min_{T \in \mc{T}_n} d(T) = \delta + \lceil \log(n-2^{\delta}) \rceil + 1.
\]
\end{lemma}

\begin{IEEEproof}
See Appendix \ref{appendix: d_n^min}.
\end{IEEEproof}

\begin{theorem}\label{theorem: co min latency}
Let $\delta = \lfloor \log(n/2) \rfloor$.
We have
\[
    \min_{S \in \mc{S}_n^{\text{co}}} l(S) = d_n^{\min} - 1 = \delta + \lceil \log(n-2^{\delta}) \rceil.
\]
\end{theorem}

\begin{IEEEproof}
This is the result by combining Theorem \ref{theorem: h bijective}, Lemma \ref{lemma: l(h(T)) = d(T)-1}, and Lemma \ref{lemma: d_n^min}.
\end{IEEEproof}

The proof of Lemma \ref{lemma: d_n^min} in Appendix \ref{appendix: d_n^min} also leads to the following construction for complexity-optimal structures with latency $d_n^{\min} - 1$.

\begin{construction}\label{construction: complexity optimal}
Let $T \in \mc{T}_n$ be a T-tree, in which there exists an edge $(a, b) \in T$ such that $D(a, b, T)$ and $D(b, a, T)$ are two complete DBTs with leaves $\{x_1, x_2, \ldots, x_{2^{\delta}}\}$ and $\{x_{2^{\delta} + 1}, x_{2^{\delta} + 2}, \ldots, x_{n}\}$, respectively, where $\delta = \lfloor \log(n/2) \rfloor$.
Return $h(T)$ as the constructed structure.
\end{construction}

\begin{figure}[t]
\centering
\includegraphics[scale = 0.5]{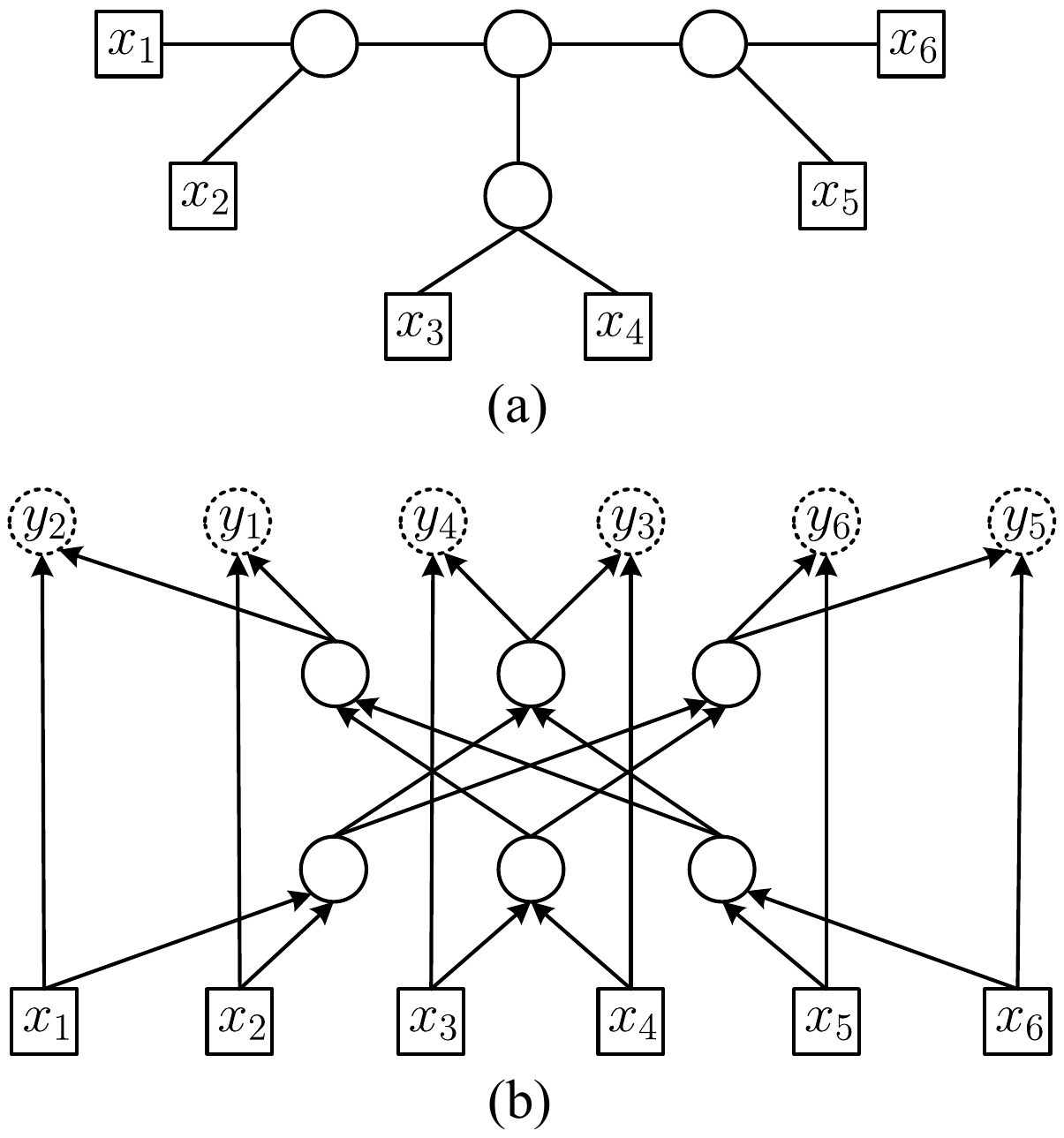}
\caption{An example of Construction \ref{construction: complexity optimal} for $n = 6$. (a) A resulted T-tree $T \in \mc{T}_6$. (b) The constructed structure $h(T) \in \mc{S}_6^{\text{co}}$ corresponding to $T$.
}
\label{fig: S6}
\end{figure}

As an example, for $n = 6$, Construction \ref{construction: complexity optimal} may lead to the T-tree $T \in \mc{T}_6$ in Fig. \ref{fig: S6}(a).
We have $d(T) = d_6^{\min} = 4$.
The constructed structure $h(T) \in \mc{S}_6^{\text{co}}$ is shown in Fig. \ref{fig: S6}(b), which has the optimal complexity $c(h(T)) = c_6^{\min} = 12$ and the minimum latency $l(h(T)) = d_6^{\min} - 1 = 3$ among $\mc{S}_6^{\text{co}}$.
It is worth mentioning that $h(T)$ in Fig. \ref{fig: S6}(b) was used in \cite{lewandowsky2018design} and \cite{lewandowsky2019design} to implement check node update for decoding regular LDPC codes with variable node degree 3 and check node degree 6.

\section{Latency-Optimal Structures}\label{section: Latency-Optimal}

In this section, we first derive the value of $l_n^{\min}$.
Then, for $n = 2^k + 1$ with $k > 0$, we propose an optimal construction for an $S \in \mc{S}_n^{\text{lo}}$ such that $c(S) = \min_{S' \in \mc{S}_n^{\text{lo}}} c(S')$.
(The construction of latency-optimal structures with other values of $n$ will be addressed later in Construction \ref{construction: general} of Section \ref{section: Tradeoff}.)

\begin{theorem}\label{theorem: l_n^min = log(n-1)}
We have $l_n^{\min} = \min_{S \in \mc{S}_n} l(S) = \lceil \log(n-1) \rceil$.
\end{theorem}

\begin{IEEEproof}
For any $S \in \mc{S}_n$ and $j \in [n]$, $E(y_j, S)$ is a DBT with leaves $X \setminus \{{x_j}\}$.
As a result, the minimum height of $E(y_j, S)$ is $\lceil \log(n-1) \rceil$ which is achievable when $E(y_j, S)$ is a complete DBT.
Since $S = \vee_{j \in [n]} E(y_j, S)$, we have $l(S) \geq \lceil \log(n-1) \rceil$, where the equality holds when each $E(y_j, S)$ is a complete DBT.
This completes the proof.
\end{IEEEproof}

According to Theorems \ref{theorem: c_n^min = 3n - 6} and \ref{theorem: l_n^min = log(n-1)}, the structure $S$ in Fig. \ref{fig: S6}(b) is both complexity-optimal and latency-optimal, i.e., $S \in \mc{S}_6^{\text{co}} \cap \mc{S}_6^{\text{lo}}$.
However, structures that are both complexity-optimal and latency-optimal rarely exist.
In fact, according to Theorems \ref{theorem: co min latency} and \ref{theorem: l_n^min = log(n-1)}, we can easily derive the following corollary.

\begin{corollary}
For $n \geq 3$, we have $\mc{S}_n^{\text{co}} \cap \mc{S}_n^{\text{lo}} \neq \emptyset$ iff $n = 3, 4, 6$.
\end{corollary}

We now propose a simple construction for latency-optimal structures when $n-1$ is a power of two.

\begin{construction}[For $n = 2^k + 1$ with $k > 0$]\label{construction: latency optimal = 2^k+1}
Let $S = (\{v_{0, j}: j \in [n]\}, \emptyset)$ with $v_{0, j} = x_j$.
For $i = 1, 2, \ldots, k$ and $j \in [n]$, create a new node $v_{i, j} \notin S$, and let $S = S + (v_{i-1, j}, v_{i, j}) + (v_{i-1, j+2^{i-1}}, v_{i, j})$, where $v_{i-1, j+2^{i-1}} = v_{i-1, j+2^{i-1}-n}$ if $j+2^{i-1} > n$.
Return $S$ as the constructed structure.
\end{construction}

\begin{theorem}\label{theorem: latency optimal = 2^k+1}
Assume $n = 2^k + 1$ with $k > 0$.
$S$ returned by Construction \ref{construction: latency optimal = 2^k+1} belongs to $\mc{S}_n^{\text{lo}}$, and we have
\[
    c(S) = n\log(n-1) = nk = \min_{S' \in \mc{S}_n^{\text{lo}}} c(S').
\]
\end{theorem}

\begin{IEEEproof}
See Appendix \ref{appendix: latency optimal = 2^k+1}.
\end{IEEEproof}

\begin{figure}[t]
\centering
\includegraphics[scale = 0.5]{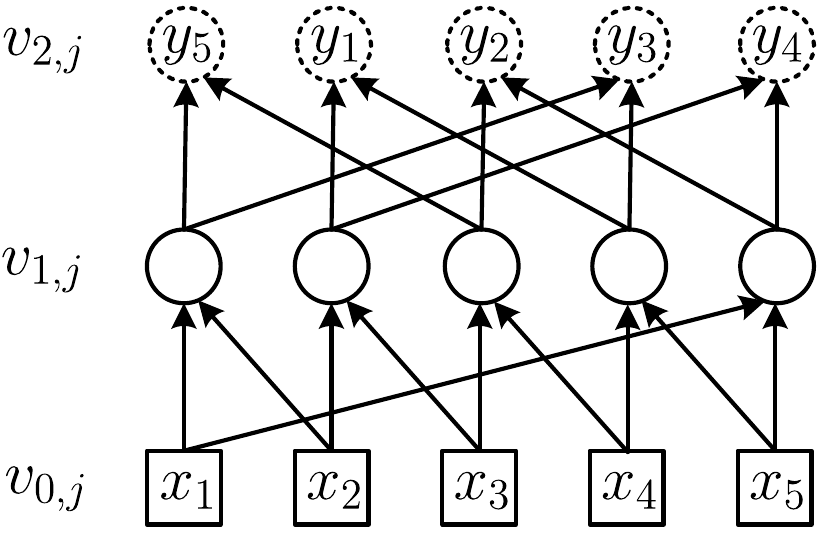}
\caption{An example of Construction \ref{construction: latency optimal = 2^k+1} for $n = 5$.
}
\label{fig: S5}
\end{figure}

Note that Construction \ref{construction: latency optimal = 2^k+1} is deterministic, i.e., the result of Construction \ref{construction: latency optimal = 2^k+1} is unique for any $n = 2^k + 1$ with $k > 0$.
An example of Construction \ref{construction: latency optimal = 2^k+1} for $n = 5$ is shown in Fig. \ref{fig: S5}, which has latency 2 and complexity 10.

\section{Tradeoff between Complexity and Latency}\label{section: Tradeoff}

A general problem is to find the minimum complexity of structures in $\mc{S}_n$ that have latencies at most $\tau$ for any given $(n, \tau)$.
We give a solution to this problem in this section.

For $n = 2$,  outgoing messages are given by $y_1 = x_2$ and $y_2 = x_1$.
Accordingly, the graph that only consists of nodes $\{x_1, x_2\}$, say $G$, can be considered as a valid (and the only) structure used for computing $\mb{y}$ for $n = 2$.
Moreover, $G$ is both complexity-optimal and latency-optimal.
For convenience, we let $\mc{S}_2 = \{G\}$.

For any $S \in \mc{S}_n$, recall that $P(S)$ is the set of all complement pairs of $S$.
Note that we must have $P(S) \neq \emptyset$.
Let $\pi(S) = \min_{\langle a, b \rangle \in P(S)} \pi(a, b)$ and ${P}_{\pi}(S) = \{\langle a, b \rangle \in P(S): \pi(a, b) = \pi(S)\}$, where $\pi(a, b)$ is equal to one plus the maximum height of $E(a, S)$ and $E(b, S)$.
As a result, we have $\pi(S) \geq \lceil \log(n) \rceil$.

\begin{lemma}\label{lemma: pi(S)}
For $n \geq 3$, we have $\pi(S) = \lceil \log(n) \rceil$ with $S$ returned by Construction \ref{construction: complexity optimal}.
\end{lemma}

\begin{IEEEproof}
Let $S$ be returned by Construction \ref{construction: complexity optimal} and let $T = h^{-1}(S)$, where $h^{-1}$ is the inverse function of $h$ defined in Theorem \ref{theorem: h bijective}.
We have $d(T) = d_n^{\min} = l(S) + 1 = \delta + \lceil \log(n-2^{\delta}) \rceil + 1$ with $\delta = \lfloor \log(n/2) \rfloor$.
There exist two leaves, say $x_i, x_j \in T$, such that the distance between $x_i$ and $x_j$ is $d(T)$.
Moreover, given $x_i$ and $x_j$, there exists a unique node $a$ (resp. $b$) such that $a$ (resp. $b$) is contained in the path from $x_i$ to $x_j$ and the distance between $x_i$ and $a$ (resp. $b$) is $\lfloor d(T)/2 \rfloor$ (resp. $\lfloor d(T)/2 \rfloor + 1$).
Note that $(a, b) \in T$.
As a result, there exists a complement pair $\langle a', b' \rangle \in P(S)$ such that $D(a, b, T) = E(a', S)$ and $D(b, a, T) = E(b', S)$.
Accordingly, the heights of $E(a', S)$ and $E(b', S)$ are $\lfloor d(T)/2 \rfloor$ and $d(T) - \lfloor d(T)/2 \rfloor - 1$, respectively.
We then have $\pi(a', b') = 1 + \max\{\lfloor d(T)/2 \rfloor, d(T) - \lfloor d(T)/2 \rfloor - 1\} = 1 + \lfloor d(T)/2 \rfloor$.
Note that $2^{\delta+1} \leq n < 2^{\delta+2}$.
If $n = 2^{\delta+1}$, we have $\pi(a', b') = 1 + \lfloor d(T)/2 \rfloor = 1 + \lfloor (\delta + \lceil \log(n-2^{\delta}) \rceil + 1) / 2 \rfloor = 1 + \delta = \lceil \log(n) \rceil$.
If $2^{\delta+1} < n < 2^{\delta+2}$, we have $\pi(a', b') = 1 + \delta + 1 = \lceil \log(n) \rceil$.
As $\pi(S) \geq \lceil \log(n) \rceil$, we finally have $\pi(S) = \lceil \log(n) \rceil$.
\end{IEEEproof}

For two integers $i$ and $j$, let $[i, j] = \{i, i +1, \ldots, j\}$, where $[i, j] = \emptyset$ if $i > j$.
We now propose a method to construct larger (in terms of input size) structures based on smaller structures.

\begin{construction}\label{construction: small2large}
If there exist $(m, n_0, \ldots, n_m)$ such that $m \in [n-1]$, $n_0 \geq m$, and $\sum_{i \in[0, m]} n_i = n + m$ with $n_i \in [2, n-1]$, we can construct an $S \in \mc{S}_n$ from any $S_i \in \mc{S}_{n_i}, \forall i \in [0, m]$ with the following steps.
\begin{enumerate}[(D1)]
\item   For each $i \in [0, m]$ and $j \in [n_i]$, refer to $x_j \in S_i$ and $y_j \in S_i$ by $a_{i, j}$ and $b_{i, j}$, respectively. (Note that if $n_i = 2$, we have $a_{i, 1} = b_{i, 2}$ and $a_{i,2} = b_{i, 1}$.)
\item   Let $S$ be the joint graph of all $S_i, i \in [0, m]$. (Simply put all $S_i, i \in [0, m]$ together into $S$ without extra operations, such as merging nodes or edges.)
\item   For each $i \in [m]$ and an arbitrary complement pair $\langle u_i, v_i \rangle \in P_{\pi}(S_i)$, let $S = S + (u_i, a_{0, i}) + (v_i, a_{0, i})$.
\item   For each $i \in [m]$ and $j \in [n_i]$, create a new node $v_{i, j} \notin S$, and let $S = S + (b_{i, j}, v_{i, j}) + (b_{0, i}, v_{i, j})$.
\item   For the nodes in $S$, label those with no incoming edges by $x_1, x_2, \ldots, x_n$; label those with no outgoing edges by $y_1, y_2, \ldots, y_n$ such that for any $j \in [n]$, $E(y_j, S)$ has leaves $\{x_1, x_2, \ldots, x_n\} \setminus \{x_j\}$; unlabel all other nodes.
\item   Return $S$ as the constructed structure.
\end{enumerate}
\end{construction}

\begin{figure}[t]
\centering
\includegraphics[scale = 0.6]{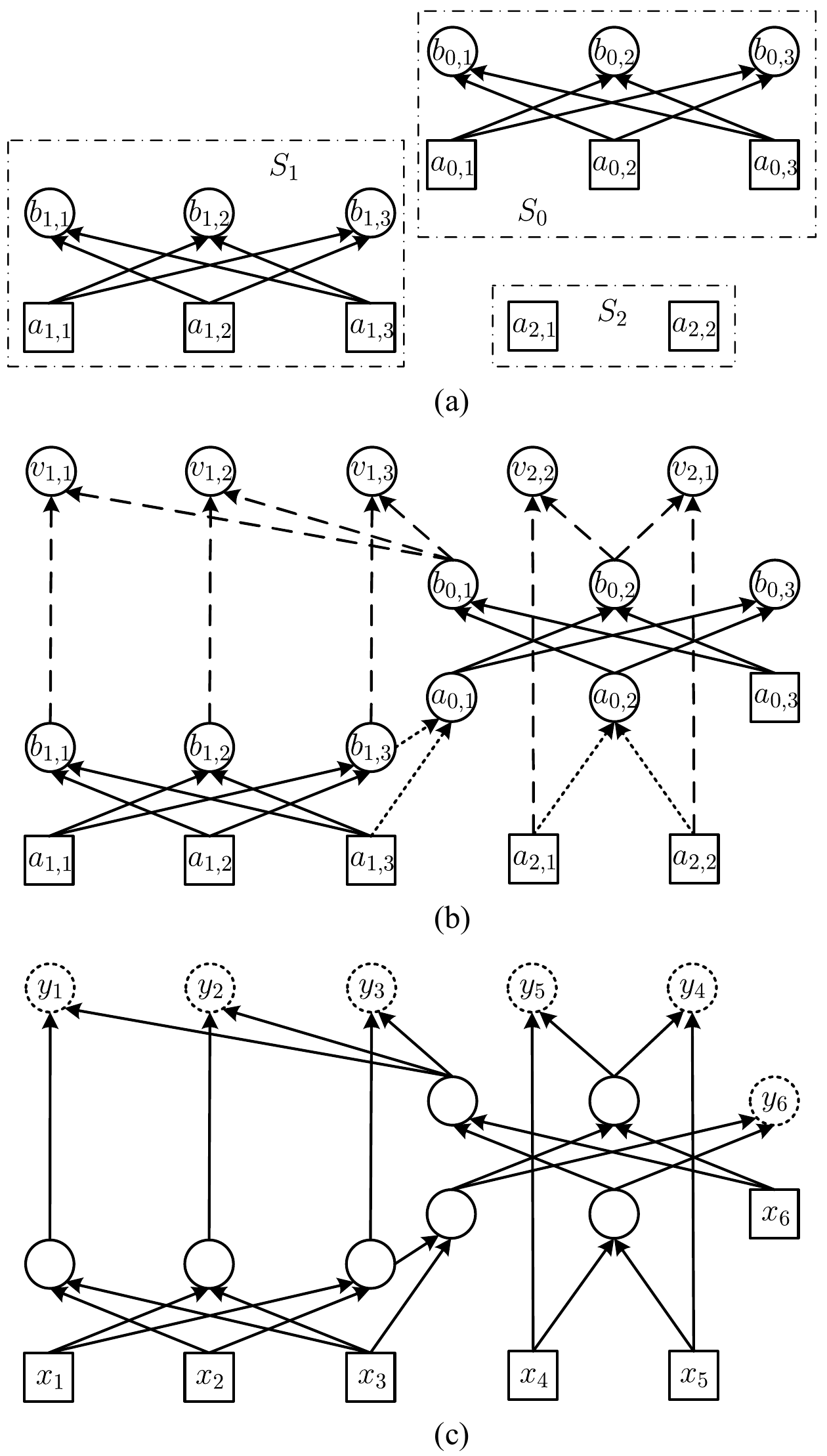}
\caption{An example of Construction \ref{construction: small2large} for $(n, m, n_0, n_1, n_2) = (6, 2, 3, 3, 2)$.
(a) $S$ formed in steps (D1) and (D2). (Note that $a_{2, 1} = b_{2, 2}$ and $a_{2,2} = b_{2, 1}$.)
(b) $S$ formed in steps (D3) and (D4), where for easy reference, the edges added into $S$ in steps (D3) and (D4) are represented by dotted and dashed arrows, respectively.
(c) $S$ formed in step (D5), which is also the constructed structure returned in step (D6).
}
\label{fig: S6_small2large}
\end{figure}

An example of Construction \ref{construction: small2large} for $(n, m, n_0, n_1, n_2) = (6, 2, 3, 3, 2)$ is shown in Fig. \ref{fig: S6_small2large}.
Moreover, we have the following result.

\begin{lemma}\label{lemma: small2large}
Use the notations in Construction \ref{construction: small2large} and let $S$ be the returned structure.
We have $S \in \mc{S}_n$.
Moreover, we have
\begin{align*}
    c(S) &= c(S_0) + \sum_{i \in [m]} \big(c(S_i) + n_i + 1\big),
    \\\pi(S) &\leq \pi(S_0) + \max_{i \in [m]}\pi(S_i), \text{~and}
    \\l(S) &\leq \max \left\{ \max_{i \in [m]}l(S_i) + 1, ~l(S_0) + 1 + \max_{i \in [m]}\pi(S_i) \right\}.
\end{align*}
\end{lemma}

\begin{IEEEproof}
Note that at the end of step (D1) (in Construction \ref{construction: small2large}), for each $i \in [0, m]$ and $j \in [n_i]$, $E(b_{i, j}, S_i)$ is a DBT with leaves $\{a_{i, j'}: j' \in [n_i]\} \setminus \{a_{i, j}\}$.
At the end of step (D3), for each $i \in [m]$, $E(a_{0, i}, S)$ is a DBT of height $\pi(S_i)$ and with leaves $\{a_{i, j'}: j' \in [n_i]\}$.
At the end of step (D4), for each $i \in [m]$ and $j \in [n_i]$, $E(v_{i, j}, S)$ is a DBT of height at most $\max \left\{ l(S_i) + 1, l(S_0) + \max_{i' \in [m]}\pi(S_{i'}) + 1 \right\}$ and with leaves $\{a_{i', j'}: i' \in [m], j' \in [n_{i'}]\} \setminus \{a_{i, j}\}$.
We can then easily verify the correctness of Lemma \ref{lemma: small2large}.
\end{IEEEproof}


%

For any non-negative integer $\tau$, let
\[
    \mc{S}_{n, \tau} = \{S \in \mc{S}_n: l(S) \leq \tau, \pi(S) = \lceil \log(n) \rceil\}.
\]
We remark that $\mc{S}_{n, \tau} \subseteq \mc{S}_{n, \tau'}$ for $\tau \leq \tau'$.
Moreover, for $n = 2$, we have $\mc{S}_{2, 0} = \mc{S}_2$.
For $n \geq 3$, we have $S \in \mc{S}_{n, d_{n}^{\min} - 1}$ if $S$ is returned by Construction \ref{construction: complexity optimal} (according to Lemma \ref{lemma: pi(S)}), and have $S \in \mc{S}_{n, \log(n-1)}$ if $S$ is returned by Construction \ref{construction: latency optimal = 2^k+1}.
Given these observations and motivated by Construction \ref{construction: small2large}, we have the following construction for a structure $S_{n, \tau} \in \mc{S}_{n, \tau}$ if $\mc{S}_{n, \tau} \neq \emptyset$; otherwise, we say that $S_{n, \tau}$ does not exist.

\begin{construction}\label{construction: general}
Let $\hat{n}$ and $\hat{\tau}$ be the maximum values of $n$ and $\tau$, respectively.
For $n = 2, 3, \ldots, \hat{n}$ and $\tau = 0, 1, \ldots, \hat{\tau}$, we construct a structure $S_{n, \tau}$ based on the following cases.
\begin{enumerate}[(E1)]
\item   If $n = 2$, let $S_{n, \tau}$ be the only structure in $\mc{S}_2$.
\item   Else if $\tau \geq d_n^{\min} - 1$, construct $S_{n, \tau}$ via Construction \ref{construction: complexity optimal}.
\item   Else if $\tau < \lceil \log(n-1) \rceil$, $S_{n, \tau}$ does not exist.
\item   Else if $2^\tau = n-1$, construct $S_{n, \tau}$ via Construction \ref{construction: latency optimal = 2^k+1}.
\item   Otherwise, let $\phi(n, \tau) = \infty$.
    For any $(m, n_0, \ldots, n_m,$ $ \tau_0, \ldots, \tau_m)$ such that
    \begin{equation}\label{eqn: condition}
        \left\{
        \begin{array}{l}
            m \in [n-1], n_0 \geq m, \\
            \sum_{i \in[0, m]} n_i = n + m, n_i \in [2, n-1],\\
            \forall i \in [0, m], S_{n_i, \tau_i} \text{~exists},\\
            \max_{i \in [m]} \lceil \log(n_i) \rceil \leq \lceil \log(n) \rceil - \lceil \log(n_0) \rceil,\\
            \max_{i \in [m]} \lceil \log(n_i) \rceil \leq \tau - 1 - \tau_0,\\
            \max_{i \in [m]} \tau_i \leq \tau - 1,\\
        \end{array}
        \right.
    \end{equation}
    construct $S$ from $S_{n_i, \tau_i}, \forall i \in [0, m]$ via Construction \ref{construction: small2large}.
    If $c(S) < \phi(n, \tau)$, let $\phi(n, \tau) = c(S)$ and $S_{n, \tau} = S$.
\end{enumerate}
\end{construction}

\begin{theorem}\label{theorem: general}
Iff $n \geq 2$ and $\tau \geq \lceil \log(n-1) \rceil$, Construction \ref{construction: general} can obtain a structure $S_{n, \tau} \in \mc{S}_{n, \tau} \neq \emptyset$.
Moreover, if $S_{n, \tau}$ exists, we have
\begin{equation}\label{eqn: c(S) upper bound}
    c(S_{n, \tau}) \leq n \lceil \log(n) \rceil - 2.
\end{equation}
\end{theorem}

\begin{IEEEproof}
See Appendix \ref{appendix: general}.
\end{IEEEproof}

We remark that in case (E5) of Construction \ref{construction: general}, we try to reuse the same subtrees (same intermediate computation results) as often as possible.
This implies that for $n \geq 2$ and $\tau \geq \lceil \log(n-1) \rceil$, we likely have
\[
    c(S_{n, \tau}) = \min_{S \in \mc{S}_{n, \tau}} c(S),
\]
which is guaranteed to be true for cases (E1), (E2), and (E4).
On the other hand, there likely exists a structure $S \in \{S' \in \mc{S}_{n}: l(S') \leq \tau\}$ such that $c(S) = \min_{S' \in \mc{S}_{n}, l(S') \leq \tau} c(S')$ and $\pi(S) = \lceil \log(n) \rceil$ (i.e., $S \in \mc{S}_{n, \tau}$).
As a result, we likely have
\begin{equation}\label{eqn: Snt is optimal}
    c(S_{n, \tau}) = \min_{S \in \mc{S}_{n}, l(S) \leq \tau} c(S),
\end{equation}
which is guaranteed to be true for cases (E1), (E2), and (E4).
However, we currently are not able to prove this result.
Formally, we give the following conjecture.

\begin{conjecture}\label{conjecture: S is optimal}
Let $S_{n, \tau}$ be returned by Construction \ref{construction: general}. Then, $S_{n, \tau}$ satisfies \eqref{eqn: Snt is optimal}.
\end{conjecture}

In general, it is not possible to enumerate $(m, n_0, \ldots, n_m,$ $ \tau_0, \ldots, \tau_m)$ by using the brute-force method in case (E5) of Construction \ref{construction: general}.
However, finding a $(m, n_0, \ldots, n_m,$ $ \tau_0, \ldots, \tau_m)$ to minimize $c(S_{n, \tau})$ is of great interest to practice.
In the rest of this section, we illustrate how to efficiently find such a $(m, n_0, \ldots, n_m,$ $ \tau_0, \ldots, \tau_m)$.

For $n \geq 2$, let $\phi(n, \tau) = c(S_{n, \tau})$, where $\phi(n, \tau) = \infty$ if $S_{n, \tau}$ does not exist.
For any $i_1, i_2 \in [0, \hat{n}], i_3 \in [0, \lceil \log(\hat{n}) \rceil]$ and $i_4 \in [0, \hat{\tau}]$, let $\eta(i_1, i_2, i_3, i_4)$ denote the minimum value of $\sum_{j\in[i_2]}(\phi(n_j, i_4)+n_j+1)$, where $n_j \in [1, 2^{i_3}]$ and $\sum_{j' \in [i_2]}n_{j'} = i_1$.
We remark that $\phi(n_j, i_4)+n_j+1$ actually corresponds to the complexity related to $S_{n_j, i_4}$ when it is used to construct a larger structure with input size $i_1$.
For example, the complexity related to $S_1$ in Fig. \ref{fig: S6_small2large} is $\phi(3, 1) + 3 + 1$.
We further remark that in (E5) of Construction \ref{construction: general}, we require $n_j \geq 2$.
However, to simplify the computation in Algorithm \ref{algo: phi}, here we allow $n_j = 1$ and define $\phi(1, \tau) = -2$ for any $\tau$ to make $\phi(1, \tau) + 1 + 1 = 0$.
Taking Fig. \ref{fig: S6_small2large} as an example, we have $(n_1, n_2, n_3) = (3, 2, 1)$, where $n_3 = 1$ is associated with the single node $a_{0,3}$.
Note that we have $\phi(n, \tau) \geq \phi(n, \tau')$ for $\tau \leq \tau'$, and we also have $\eta(i_1, i_2, i_3, i_4) \geq \eta(i_1, i_2, i'_3, i'_4)$ for $i_3 \leq i'_3$ and $i_4 \leq i'_4$.
We can compute $\phi$ and $\eta$ by using the proposed Algorithm \ref{algo: phi}.

\begin{algorithm}[t!]
\small 
    \caption{Computation of $\phi$ and $\eta$}
    \label{algo: phi}
    \begin{algorithmic}[1]
        \REQUIRE    $\hat{n}$ and $\hat{\tau}$.
        \ENSURE     $\phi$ and $\eta$.
        \STATE  For any $i_1, i_2 \in [0, \hat{n}], i_3 \in [0, \lceil \log(\hat{n}) \rceil]$ and $i_4 \in [0, \hat{\tau}]$, set $\eta(i_1, i_2, i_3, i_4)$ as $0$ if $i_1 == i_2$ and as $\infty$ otherwise.\label{code: phi @ init}

        \FOR{$n = 2, 3, \ldots, \hat{n}$ and $\tau = 0, 1, \ldots, \hat{\tau}$}
            \STATE  $//$Compute $\phi(n, \tau)$
            \IF{$n == 2$}
                \STATE  $\phi(n, \tau) = 0$.
            \ELSIF{$\tau \geq d_n^{\min} - 1$}
                \STATE  $\phi(n, \tau) = 3n - 6$.
            \ELSIF{$\tau < \lceil \log(n-1) \rceil$}
                \STATE  $\phi(n, \tau) = \infty$.
            \ELSIF{$2^\tau == n-1$}
                \STATE  $\phi(n, \tau) = n \tau$.
            \ELSE
                \STATE  $\phi(n, \tau) = \infty$.\label{code: phi @ compute phi}
                \FOR{$n_0 = 2, 3, \ldots, {n}-1$ and $\tau_0 = 0, 1, \ldots, \tau-1$}
                    \STATE  $\theta = \min\left\{ \lceil \log(n) \rceil - \lceil \log(n_0) \rceil, \tau - 1 - \tau_0 \right\}$.
                    \STATE  $\omega = \phi(n_0, \tau_0) + \eta(n, n_0, \theta, \tau-1)$.
                    \STATE  $\phi(n, \tau) = \min\{ \phi(n, \tau), \omega \}$.\label{code phi @ sol phi}
                \ENDFOR\label{code: phi @ compute phi end}
            \ENDIF

            \STATE  $//$Update $\eta$ by using $\phi(n, \tau)$
            \FOR{$i_1 = n, n+1, \ldots, \hat{n}$, $i_2 = 1, 2, \ldots, \hat{n}$, and $i_3 = \lceil \log(n) \rceil, \lceil \log(n) \rceil+1, \ldots, \lceil \log(\hat{n}) \rceil$}\label{code: phi @ update eta}
                \STATE  $\lambda = \eta(i_1-n, i_2-1, i_3, \tau) + \phi(n, \tau) + n + 1$.
                \STATE  $\eta(i_1, i_2, i_3, \tau) = \min\{\eta(i_1, i_2, i_3, \tau), \lambda\}$.\label{code phi @ sol eta}
            \ENDFOR\label{code: phi @ update eta end}
        \ENDFOR
    \end{algorithmic}
\end{algorithm}

In Algorithm \ref{algo: phi}, line \ref{code: phi @ init} is to initialize $\eta$ by using $\phi(1, \tau), \forall \tau \geq 0$.
Lines \ref{code: phi @ compute phi}--\ref{code: phi @ compute phi end} correspond to case (E5) in Construction \ref{construction: general}.
More specifically, for any $n_0 \in [2, n-1]$ and $\tau_0 \in [0, \tau-1]$ such that $\omega < \infty$, without loss of generality, assume that $\eta(n, n_0, \theta, \tau-1) = \sum_{i\in[n_0]}(\phi(n_i, \tau-1) + n_i + 1)$ with $n_1 \geq n_2 \geq \cdots \geq n_m \geq 2 > n_{m+1} = n_{m+1} = \cdots = n_{n_0} = 1$.
Then, $(m, n_0, \ldots, n_m,$ $ \tau_0, \ldots, \tau_m)$ satisfy \eqref{eqn: condition}, and we have $c(S) = \omega$ with $S$ constructed from $S_{n_i, \tau_i}, \forall i \in [0, m]$ via Construction \ref{construction: small2large}.
As a result, according to the definition of $\eta$, $\phi(n, \tau)$ computed via lines \ref{code: phi @ compute phi}--\ref{code: phi @ compute phi end} is equal to $c(S_{n, \tau})$ with $S_{n, \tau}$ given in case (E5) of Construction \ref{construction: general}.
Moreover, lines \ref{code: phi @ update eta}--\ref{code: phi @ update eta end} are to update $\eta$ by using $\phi(n, \tau)$, so as to keep $\eta(i_1, i_2, i_3, \tau)$ to be the minimum value of $\sum_{j\in[i_2]}(\phi(n_j, \tau)+n_j+1)$, where $n_j \in [1, n]$ and $\sum_{j' \in [i_2]}n_{j'} = i_1$.

\begin{theorem}
For $n \geq 2$ and $\tau \geq 0$, let $\phi(n, \tau)$ be computed via Algorithm \ref{algo: phi}, and let $S_{n, \tau}$ be returned by Construction \ref{construction: general}.
We have $\phi(n, \tau) = c(S_{n, \tau})$ if $\tau \geq \lceil \log(n-1) \rceil$, and $\phi(n, \tau) = \infty$ otherwise.
\end{theorem}

\begin{IEEEproof}
The statement is true according to the above discussions regarding Algorithm \ref{algo: phi}.
\end{IEEEproof}

Note that $\phi(n, \tau) = 3n - 6$ for $\tau \geq d_n^{\min} - 1$, where $d_n^{\min} \leq 2\log(n) + 1$ according to Lemma \ref{lemma: d_n^min}.
We only need to compute $\phi(n, \tau)$ for $\tau < d_n^{\min} - 1$.
As a result, the complexity of Algorithm \ref{algo: phi} for computing $\phi(n, \tau)$ is $O(n^3 \log^2(n))$.
For easy reference, we present $\phi(n, \tau)$ for some typical $(n, \tau)$ in Table \ref{table: phi}.
We can see that $\phi(n, \tau)$ is generally much smaller than $n \lceil \log(n) \rceil - 2$, the upper bound given by \eqref{eqn: c(S) upper bound}.

\begin{table*}[!t]
\renewcommand{\arraystretch}{1.3}
\caption{Some Typical Values and Upper Bounds (UBs)  of $\phi(n, \tau)$:
$n \geq 2, \tau = \lceil \log(n-1) \rceil + i, i = 0, 1, \ldots$;
$\phi(n, \tau)$ Is the Complexity of the Structure $S_{n, \tau}$ Obtained from Construction \ref{construction: general};
$S_{n, \tau}$ Has $n$ Input Nodes and Its Latency Is at Most $\tau$;
for $\tau < \underline{\tau} = \lceil \log(n-1) \rceil$, $S_{n, \tau}$ Does Not Exist;
for $\tau \geq \overline{\tau} = \delta + \lceil \log(n - 2^{\delta}) \rceil $ with $\delta = \lfloor \log(n/2) \rfloor$, We Have $\phi(n, \tau) = 3n - 6$;
the UB Is $n \lceil \log(n) \rceil - 2$  Given by \eqref{eqn: c(S) upper bound}.
}
\label{table: phi}
\centering
\begin{tabular}{lccccc||lcccccc}
\toprule
$n,~\underline{\tau},~\overline{\tau} \setminus i$   &0  &1      &2      &3      &UB &$n,~\underline{\tau},~\overline{\tau} \setminus i$       &0      &1      &2      &3      &4  &UB\\
\midrule
1, --, -- &--   &--     &--     &--     &--     &33, 5, 9       &165    &114    &99     &94     &93     &196\\
2, 0, 0 &0      &0      &0      &0      &0      &34, 6, 9       &118    &102    &97     &96     &96     &202\\
3, 1, 1 &3      &3      &3      &3      &4      &35, 6, 9       &122    &105    &100    &99     &99     &208\\
4, 2, 2 &6      &6      &6      &6      &6      &36, 6, 9       &126    &108    &103    &102    &102    &214\\
5, 2, 3 &10     &9      &9      &9      &13     &37, 6, 9       &133    &115    &106    &105    &105    &220\\
6, 3, 3 &12     &12     &12     &12     &16     &38, 6, 9       &137    &118    &109    &108    &108    &226\\
7, 3, 4 &18     &15     &15     &15     &19     &39, 6, 9       &141    &122    &112    &111    &111    &232\\
8, 3, 4 &22     &18     &18     &18     &22     &40, 6, 9       &145    &125    &115    &114    &114    &238\\
9, 3, 5 &27     &22     &21     &21     &34     &41, 6, 9       &159    &132    &122    &117    &117    &244\\
10, 4, 5        &25     &24     &24     &24     &38     &42, 6, 9       &163    &135    &125    &120    &120    &250\\
11, 4, 5        &32     &27     &27     &27     &42     &43, 6, 9       &168    &139    &128    &123    &123    &256\\
12, 4, 5        &36     &30     &30     &30     &46     &44, 6, 9       &172    &142    &131    &126    &126    &262\\
13, 4, 6        &45     &36     &33     &33     &50     &45, 6, 9       &179    &146    &135    &129    &129    &268\\
14, 4, 6        &50     &39     &36     &36     &54     &46, 6, 9       &183    &149    &138    &132    &132    &274\\
15, 4, 6        &57     &43     &39     &39     &58     &47, 6, 9       &188    &153    &141    &135    &135    &280\\
16, 4, 6        &62     &46     &42     &42     &62     &48, 6, 9       &192    &156    &144    &138    &138    &286\\
17, 4, 7        &68     &51     &46     &45     &83     &49, 6, 10      &243    &176    &153    &144    &141    &292\\
18, 5, 7        &54     &49     &48     &48     &88     &50, 6, 10      &250    &180    &156    &147    &144    &298\\
19, 5, 7        &61     &52     &51     &51     &93     &51, 6, 10      &259    &184    &159    &150    &147    &304\\
20, 5, 7        &65     &55     &54     &54     &98     &52, 6, 10      &266    &188    &162    &153    &150    &310\\
21, 5, 7        &72     &62     &57     &57     &103    &53, 6, 10      &277    &192    &167    &156    &153    &316\\
22, 5, 7        &76     &65     &60     &60     &108    &54, 6, 10      &284    &196    &170    &159    &156    &322\\
23, 5, 7        &80     &69     &63     &63     &113    &55, 6, 10      &293    &200    &173    &162    &159    &328\\
24, 5, 7        &84     &72     &66     &66     &118    &56, 6, 10      &300    &204    &176    &165    &162    &334\\
25, 5, 8        &108    &81     &72     &69     &123    &57, 6, 10      &325    &210    &182    &169    &165    &340\\
26, 5, 8        &114    &84     &75     &72     &128    &58, 6, 10      &332    &214    &185    &172    &168    &346\\
27, 5, 8        &122    &89     &78     &75     &133    &59, 6, 10      &341    &218    &189    &175    &171    &352\\
28, 5, 8        &128    &92     &81     &78     &138    &60, 6, 10      &348    &222    &192    &178    &174    &358\\
29, 5, 8        &138    &98     &85     &81     &143    &61, 6, 10      &359    &226    &196    &181    &177    &364\\
30, 5, 8        &144    &102    &88     &84     &148    &62, 6, 10      &366    &230    &199    &184    &180    &370\\
31, 5, 8        &152    &106    &91     &87     &153    &63, 6, 10      &375    &234    &203    &187    &183    &376\\
32, 5, 8        &158    &110    &94     &90     &158    &64, 6, 10      &382    &238    &206    &190    &186    &382\\
\bottomrule
\end{tabular}
\end{table*}

To find a $(m, n_0, \ldots, n_m,$ $ \tau_0, \ldots, \tau_m)$ to minimize $c(S_{n, \tau})$ in case (E5) of Construction \ref{construction: general}, we only need to record the solutions to $\phi(n, \tau)$ and $\eta(i_1, i_2, i_3, \tau)$ in lines \ref{code phi @ sol phi} and \ref{code phi @ sol eta} of Algorithm \ref{algo: phi}, respectively.
More specifically, record $(n_0, \tau_0)$ such that $\phi(n, \tau) = \phi(n_0, \tau_0) + \eta(n, n_0, \theta, \tau-1)$, and record $(n, i_3)$ such that $\eta(i_1, i_2, i_3, \tau) = \eta(i_1-n, i_2-1, i_3, \tau) + \phi(n, \tau) + n + 1$.
In this case, we can find a $(m, n_0, \ldots, n_m,$ $ \tau_0, \ldots, \tau_m)$ by traceback.

We remark that for the computation at a variable node of LDPC codes, there exists a unique incoming message, say $x_1$, which corresponds to the received channel message.
In this case, $y_1$ is used for hard decision of the corresponding transmitted bit and should be computed from $\mb{x}$ without excluding $x_1$.
We can slightly modify the structure $S_{n, \tau}$ returned by Construction \ref{construction: general} to perfectly match the aforementioned variable node computation.
More specifically, for an arbitrary complement pair $\langle a, b \rangle \in P_{\pi}(S_{n, \tau})$, let $S = S_{n, \tau} - y_1 + (a, y_1) + (b, y_1)$.
Then, $E(y_1, S)$ is a DBT of height $\lceil \log(n) \rceil$ and with leaves $X = \{x_1, x_2, \ldots, x_n\}$, and $E(y_j, S) = E(y_j, S_{n, \tau}), \forall j = 2, 3, \ldots, n$.
This indicates that $S$ can be used to implement the aforementioned variable node computation. Moreover, we have $c(S) = c(S_{n, \tau})$, and $l(S) = l(S_{n, \tau}) + 1$ if $n = 2^{\tau} + 1$; otherwise, $l(S) = l(S_{n, \tau})$.

\section{Conclusion}\label{section: Conclusion}

Let $S \in \mc{S}_n$ be an arbitrary structure satisfying Definition \ref{definition: structure y}.
First, we have proved that the minimum complexity of $S$ is $3n - 6$, and if $S$ has such complexity, its minimum latency is $\delta + \lceil \log(n-2^{\delta}) \rceil$ with $\delta = \lfloor \log(n/2) \rfloor$.
Next, we have proved that the minimum latency of $S$ is $\lceil \log(n-1) \rceil$, and if $S$ has such latency, its minimum complexity is $n \log(n-1)$ for $n = 2^k + 1$ with $k > 0$.
Finally, given $(n, \tau)$ with $\tau \geq \lceil \log(n-1) \rceil$, we have proposed a construction, i.e., Construction \ref{construction: general}, for a structure $S_{n, \tau}$  which we conjecture to have the minimum complexity among structures with latencies at most $\tau$.
Construction \ref{construction: general} can run in $O(n^3 \log^2(n))$ time, and the obtained $S_{n, \tau}$ has complexity at most (generally much smaller than) $n \lceil \log(n) \rceil - 2$.
One left problem is to verify whether $S_{n, \tau}$ returned by Construction \ref{construction: general} achieves the minimum complexity among structures with latencies at most $\tau$, i.e., to prove/disprove Conjecture \ref{conjecture: S is optimal}.

\appendices

\section{Proof of Theorem \ref{theorem: Sn property}}\label{appendix: Sn property}

For $n = 3$, the only structure in $\mc{S}_n^{\text{co}}$, as shown in Fig. \ref{fig: S3}(a), fulfills properties (C1)--(C5).
Assume that for $n = k - 1 \geq 3$, structures in $\mc{S}_{k-1}^{\text{co}}$ fulfill properties (C1)--(C5).
We now prove for $n = k$,  structures in $\mc{S}_{k}^{\text{co}}$ also fulfill properties (C1)--(C5).
Let $S \in \mc{S}_{k}^{\text{co}}$ be an arbitrary structure.

\emph{Proof of (C1):} $y_k$ has exactly two incoming edges in $S$, say $(a_1, y_k), (b_1, y_k) \in S$.
Moreover, as discussed earlier, $x_k$ has exactly two outgoing edges in $S$, say $(x_k, a_2), (x_k, b_2) \in S$, and we have $f(S) \in \mc{S}_{k-1}^{\text{co}}$ such that $f(S)$ fulfills properties (C1)--(C3).
As a result, nodes $y_k, x_k, a_1, b_1, a_2, b_2$ and their edges must form a subgraph of $S$ exactly the same as that in Fig. \ref{fig: xn_Sn}(a) (with $n = k$), and this subgraph changes to a subgraph in $f(S)$ exactly the same as that in Fig. \ref{fig: xn_Sn}(d).
Note that we have $S - y_k - x_k - a_1 - b_1 - a_2 - b_2 = f(S) - a - b$.
Hence, $S$ fulfills property (C1).

\emph{Proof of (C2):} Note that in the following proof, the definitions of $(a_1, b_1, a_2, b_2)$ are from inside the proof of (C1) and Fig. \ref{fig: xn_Sn}(a).
We first have $E(a_1, S) = E(a, f(S))$, $E(b_1, S) = E(b, f(S))$, $L(a_2, S) = L(a, f(S))$, and $L(b_2, S) = L(b, f(S))$.
As a result, we have $\langle a, b \rangle \in P(f(S))$, since $X\setminus\{x_k\} = \{x_1, x_2, \ldots, x_{k-1}\} \subseteq E(y_k, S)$.
This indicates that $E(a, f(S)) - a, E(b, f(S)) - b, L(a, f(S)) - a$, and $L(b, f(S)) - b$ pairwise do not share the same node in $f(S)$, and we have $Y\setminus \{y_k\} = \{y_1, y_2, \ldots, y_{k-1}\} \subseteq L(a, f(S)) \cup L(b, f(S))$ since $f(S)$ fulfills property (C2).
On the other hand, $E(a, f(S))$ and $E(b, f(S))$ are two DBTs. Meanwhile, since $f(S)$ fulfills property (C1), the undirected versions of $L(a, f(S))$ and $L(b, f(S))$ are two binary trees rooted at $a$ and $b$ in $f(S)$, respectively.
Therefore, $E(a, f(S)) - a, E(b, f(S)) - b, L(a, f(S)) - a$, and $L(b, f(S)) - b$ contain $4(k-1) - 8$ nodes, which are exactly all the nodes in $f(S) - a - b$.
Accordingly, $E(y_k, S)$ and $L(x_k, S)$ contain $4k - 6$ nodes, which are exactly all the nodes in $S$.

For any $v \in E(y_k, S)$, if $v = y_k$, we obviously have $x_j \in E(v, S) \iff y_j \notin L(v, S), \forall j \in [k]$.
Assume $v \neq y_k$.
As discussed above, there exists a unique $v' \in E(a, f(S)) \cup E(b, f(S))$ such that $E(v', f(S)) = E(v, S)$.
Meanwhile, we have $y_j \in L(v', f(S)) \iff y_j \in L(v, S), \forall j \in [k-1]$.
Since $f(S)$ fulfills property (C2), we have $x_j \in E(v', f(S)) \iff y_j \notin L(v', f(S)), \forall j \in [k-1]$.
Therefore, we have $x_j \in E(v, S) \iff y_j \notin L(v, S), \forall j \in [k]$ by further noting that $x_k \notin E(v, S)$ and $y_k \in L(v, S)$.

On the other hand, for any $v \in L(x_k, S)$, if $v = x_k$, we obviously have $x_j \in E(v, S) \iff y_j \notin L(v, S), \forall j \in [k]$.
For $v \neq x_k$, let $v' \in L(a, f(S)) \cup L(b, f(S))$ such that $L(v', f(S)) = L(v, S)$.
We can similarly derive $x_j \in E(v, S) \iff y_j \notin L(v, S), \forall j \in [k]$.
As a result, $S$ fulfills property (C2).

\emph{Proof of (C3):} For any $v \in E(y_k, S)$, if $v = y_k$, we obviously have $\langle v, \bar{v} \rangle \in P(S) \iff \bar{v} = x_k$.
Assume $v \neq y_k$.
There exists a unique $v' \in E(a, f(S)) \cup E(b, f(S))$ such that $E(v', f(S)) = E(v, S)$.
Since $f(S)$ fulfills property (C3), there exists a unique $\bar{v}' \in f(S)$ such that $\langle v', \bar{v}'\rangle  \in P(f(S))$.
On the one hand, we must have $\bar{v}' \in L(a, f(S)) \cup L(b, f(S))$.
As a result, there exists a unique $\bar{v} \in L(x_k, S)$ such that $L(\bar{v}, S) = L(\bar{v}', f(S))$.
On the other hand, since $f(S)$ fulfills property (C2), we have for $j \in [k-1]$, $x_j \in E(v', f(S)) \iff x_j \notin E(\bar{v}', f(S)) \iff y_j \in L(\bar{v}', f(S))$.
Therefore, we have for $j \in [k]$, $x_j \in E(v, S) \iff y_j \in L(\bar{v}, S) \iff x_j \notin E(\bar{v}, S)$  by further noting that $x_k \notin E(v, S)$, $y_k \notin L(\bar{v}, S)$ and $S$ fulfills property (C2).
This indicates that $\langle v, \bar{v}\rangle  \in P(S)$.
Note that  $E(y_k, S)$ contains half nodes of $S$, $L(x_k, S)$ contains another half nodes of $S$,  and each $v \in E(y_k, S)$ leads to a unique $\bar{v} \in L(x_k, S)$ such that $\langle v, \bar{v}\rangle  \in P(S)$.
Hence, $S$ fulfills property (C3), and we also have $|P(S)| = 2k - 3$.

\emph{Proof of (C4):}
On the one hand, for any $S' \in \mc{S}_{k-1}^{\text{co}}$ and $\langle a, b \rangle \in P(S')$, we have $g(a, b, S') \in \mc{S}_{k}^{\text{co}}$.
This implies $\{g(a, b, S'): S' \in \mc{S}_{k-1}^{\text{co}}, \langle a, b \rangle \in P(S')\} \subseteq \mc{S}_{k}^{\text{co}}$.
On the other hand, for any $S \in \mc{S}_{k}^{\text{co}}$, we have $f(S) \in \mc{S}_{k-1}^{\text{co}}$, and there exists $\langle a, b \rangle \in P(f(S))$ such that $S = g(a, b, f(S))$.
This implies $\mc{S}_{k}^{\text{co}} \subseteq \{g(a, b, S'): S' \in \mc{S}_{k-1}^{\text{co}}, \langle a, b \rangle \in P(S')\}$.
As a result, we have $\mc{S}_{k}^{\text{co}} = \{g(a, b, S'): S' \in \mc{S}_{k-1}^{\text{co}}, \langle a, b \rangle \in P(S')\}$, indicating that $\mc{S}_{k}^{\text{co}}$ fulfills property (C4).

\emph{Proof of (C5):}
For any $S' \in \mc{S}_{k-1}^{\text{co}}$, on the one hand, we have $g(a, b, S') \neq g(a', b', S')$ for any $\langle a, b \rangle, \langle a', b' \rangle \in P(S')$ with $\langle a, b \rangle \neq \langle a', b' \rangle$.
This implies $|\{g(a, b, S'): \langle a, b \rangle \in P(S')\}| = |P(S')| = 2(k-1)-3$, since $S'$ fulfills property (C3).
On the other hand, for any $S' \neq S'' \in \mc{S}_{k-1}^{\text{co}}$, we have $\{g(a, b, S'): \langle a, b \rangle \in P(S')\} \cap \{g(a, b, S''): \langle a, b \rangle \in P(S'')\} = \emptyset$.
Therefore, we have $|\mc{S}_{k}^{\text{co}}| = |\{g(a, b, S'): S' \in \mc{S}_{k-1}^{\text{co}}, \langle a, b \rangle \in P(S')\}| = |\mc{S}_{k-1}^{\text{co}}| \cdot (2(k-1)-3) = (2k - 5)!!$, indicating that $\mc{S}_{k}^{\text{co}}$ fulfills property (C5).

\section{Proof of Theorem \ref{theorem: h bijective}}\label{appendix: h bijective}

For any $T \in \mc{T}_n$ and $j \in [n]$, $x_j$ is a leaf in $T$.
Let $(a, x_j) \in T$ be the only edge of $x_j$.
$D(a, x_j, T)$ is a DBT with root $y_j$ and leaves $X \setminus \{x_j\}$.
As a result, we have $h(T) \in \mc{S}_n$.
On the other hand, we have $E(h(T)) = \cup_{j \in [n], (a, x_j) \in T} E(D(a, x_j, T)) = D(T)$, leading to $c(h(T)) = |E(h(T))| - n = |D(T)| - n = 3n - 6$.
Therefore, we have $h(T) \in \mc{S}_n^{\text{co}}$.
Moreover, for any $T \neq T' \in \mc{T}_n$, we obviously have $h(T) \neq h(T')$, indicating that $h$ is an injection from $\mc{T}_n$ to $\mc{S}_n^{\text{co}}$.
In the following, we prove $|\mc{T}_n| = |\mc{S}_n^{\text{co}}| = (2n - 5)!!$ such that $h$ is surjective and the proof is completed.

Assume $n \geq 4$.
For any $T' \in \mc{T}_{n-1}$ and $(a, b) \in T'$, let $\beta(a, b, T') = T' - (a, b) + (x_n, v) + (a, v) + (b, v)$, where $v$ is a new internal node (unlabelled) added into $T'$.
We have $\beta(a, b, T') \in \mc{T}_n$.
This implies $\{\beta(a, b, T'): T' \in \mc{T}_{n-1}, (a, b) \in T'\} \subseteq \mc{T}_n$.
On the other hand, for any $T \in \mc{T}_n$, let $\alpha(T) = T - x_n - v + (a, b)$, where $v, a, b$ fulfill $(x_n, v), (a, v), (b, v) \in T$ (note that $v$ and $(a, b)$ are unique given $T$).
We have $\alpha(T) \in \mc{T}_{n-1}$ and $T = \beta(a, b, \alpha(T))$.
This implies $\mc{T}_n \subseteq \{\beta(a, b, T'): T' \in \mc{T}_{n-1}, (a, b) \in T'\}$.
As a result, we have $\mc{T}_n = \{\beta(a, b, T'): T' \in \mc{T}_{n-1}, (a, b) \in T'\}$.

Moreover, note that for any $T' \in \mc{T}_{n-1}$, we have $\beta(a, b, T') \neq \beta(a', b', T')$ for $(a, b), (a', b') \in T'$ with $(a, b) \neq (a', b')$.
This implies $|\{\beta(a, b, T'): (a, b) \in T'\}| = 2(n-1)-3$.
Meanwhile, we have $\{\beta(a, b, T'): (a, b) \in T'\} \cap \{\beta(a, b, T''): (a, b) \in T''\} = \emptyset$ for any $T' \neq T'' \in \mc{T}_{n-1}$.
As a  result, we have $|\mc{T}_n| = (2(n-1) - 3) \cdot |\mc{T}_{n-1}| = (2(n-1) - 3) \cdot (2(n-2) - 3) \cdot |\mc{T}_{n-2}|  = (2n - 5)!!$, since $|\mc{T}_3| = 1 = (2\cdot3 - 5)!!$.
This completes the proof.

\section{Proof of Lemma \ref{lemma: d_n^min}}\label{appendix: d_n^min}

Let $\delta = \lfloor \log(n/2) \rfloor$.
We have $2^{\delta + 1} \leq n < 2^{\delta + 2}$.
Let $T_1$ and $T_2$ be two complete binary trees with leaves $\{x_1, x_2, \ldots, x_{2^{\delta}}\}$ and $\{x_{2^{\delta} + 1}, x_{2^{\delta} + 2}, \ldots, x_{n}\}$, respectively.
Since $T_1$ is a perfect binary tree, $T_1$ has height $h_1 = \delta$ and diameter $d(T_1) = 2 \delta$.
Meanwhile, $T_2$ has height $h_2 = \lceil \log(n-2^{\delta}) \rceil \leq \delta + 2$, since $n-2^{\delta} < 3 \cdot 2^{\delta}$.
More specifically, the left subtree of $T_2$ has height $h_2 - 1$, and the right subtree of $T_2$ has height at most $\delta$, leading to $d(T_2) \leq \max\{ 2(h_2 - 1), 2 \delta, h_2 - 1 + \delta + 2 \} = h_2 + \delta + 1$.
Furthermore, there exists a $T \in \mc{T}_n$ and an edge $(a, b) \in T$ such that $D(a, b, T)$ and $D(b, a, T)$ are the directed versions of $T_1$ and $T_2$, respectively.
We have $d_n^{\min} \leq d(T) = \max\{d(T_1), d(T_2), h_1 + h_2 + 1\} = h_1 + h_2 + 1 = \delta + \lceil \log(n-2^{\delta}) \rceil + 1$.

On the other hand, for any $T \in \mc{T}_n$, assume that the distance between $x_i$ and $x_j$ is equal to $d(T)$.
The height of $D(v, x_i, T)$ with $(v, x_i) \in T$ is $d(T) - 1$ such that $D(v, x_i, T)$ contains at most $2^{d(T)-1}$ leaves.
We must have $2^{d(T)-1} \geq n - 1$, leading to $d(T) \geq \lceil \log(n-1)\rceil + 1 \geq \delta + 1$.
As a result, there exists a unique node $a$ (resp. $b$) such that $a$ (resp. $b$) is contained in the path from $x_i$ to $x_j$ and the distance between $x_i$ and $a$ (resp. $b$) is $\delta$ (resp. $\delta + 1$).
Note that we have $(a, b) \in T$.
The height of $D(a, b, T)$ is $\delta$ and hence $D(a, b, T)$ contains at most $2^{\delta}$ leaves.
Meanwhile, the height of $D(b, a, T)$ is $d(T) - \delta - 1$ and hence $D(b, a, T)$ contains at most $2^{d(T) - \delta - 1}$ leaves.
Therefore, we must have $2^{\delta} + 2^{d(T) - \delta - 1} \geq n$, leading to $d(T) \geq \delta + \lceil \log(n-2^{\delta}) \rceil + 1$.
This implies $d_n^{\min} \geq \delta + \lceil \log(n-2^{\delta}) \rceil + 1$.
Combining with the previous result $d_n^{\min} \leq  \delta + \lceil \log(n-2^{\delta}) \rceil + 1$, the proof is completed.

\section{Proof of Theorem \ref{theorem: latency optimal = 2^k+1}}\label{appendix: latency optimal = 2^k+1}

Use the notations in Construction \ref{construction: latency optimal = 2^k+1} and let $S$ be the returned structure.
For each $j \in [n]$, $E(v_{k, j}, S)$ is a perfect DBT of height $k$ and with leaves $X \setminus \{x_{j-1}\}$, where we let $x_0 = x_n$.
This indicates  $y_j = v_{k, j}$ and $S \in \mc{S}_n^{\text{lo}}$.
On the other hand, the computation nodes in $S$ are $\{v_{i, j}: i\in[k], j\in[n]\}$.
We thus have $c(S) = n\log(n-1) = nk$.
We are now to prove $\min_{S' \in \mc{S}_n^{\text{lo}}} c(S') = nk$.

Given an arbitrary structure $S' \in \mc{S}_n^{\text{lo}}$.
For any $j \in [n]$, $E(y_j, S')$ must be a perfect DBT of height $k$.
This also implies that for any $a \in S'$, $E(a, S')$ is a perfect DBT.
For $i \in [k]$, let $A_i = \{a \in S': E(a, S') \text{~has height $i$}\}$.
Accordingly, we have $A_k = Y = \{y_1, y_2, \ldots, y_n\}$.
Our idea is to prove $|A_i| \geq n, \forall i \in [k]$ such that $c(S') = \sum_{i \in [k]} |A_i| \geq n k$, which can complete the proof.

For any $a \in S'$, let $\Gamma(a) = (\gamma_1, \gamma_2, \ldots, \gamma_{n})$, where for any $j \in [n]$, $\gamma_j = 1$ if $x_j \in E(a, S')$ and $\gamma_j = 0$ otherwise.
Meanwhile, for any $i \in [k]$ and $A \subseteq A_i$, let $\Gamma(A) = \oplus_{a \in A} \gamma(a)$, where $\oplus$ is the component-wise XOR operation.
If $A = \emptyset$, let $\Gamma(A) = (0, 0, \ldots, 0)$ ($n$ zeros in total).
Moreover, let $\Gamma(i) = \{\Gamma(A): A \subseteq A_i, |A| \text{~is even}\}, \forall i \in [k]$.

On the one hand, we have $\Gamma(a) = \Gamma(a_1) \oplus \Gamma(a_2)$ for any $(a_1, a), (a_2, a) \in S'$.
This leads to $\Gamma(k) \subseteq \Gamma(k-1) \subseteq \cdots \subseteq \Gamma(1)$.
On the other hand, for any $A \subseteq A_k = Y$ with even $|A|$, we have $\Gamma(A) = (\gamma_1, \gamma_2, \ldots, \gamma_{n})$, where for any $j \in [n]$, $\gamma_j = 1$ if $y_j \in A$ and $\gamma_j = 0$ otherwise.
This leads to $|\Gamma(k)| = 2^{n-1}$.
As a result, we have $|\Gamma(i)| \geq 2^{n-1}, \forall i \in [k]$, indicating that $|A_i| \geq n$.
This completes the proof.

\section{Proof of Theorem \ref{theorem: general}}\label{appendix: general}

First of all, we have $\mc{S}_{n, \tau} \neq \emptyset$ iff $n \geq 2$ and $\tau \geq \lceil \log(n-1) \rceil$.
Assume $n \geq 2$ and $\tau \geq \lceil \log(n-1) \rceil$.
If $(n, \tau)$ fulfill case (E1) (in Construction \ref{construction: general}), we have $S_{n, \tau} \in \mc{S}_{n, 0} \subseteq \mc{S}_{n, \tau}$ and \eqref{eqn: c(S) upper bound} holds.
Else if $(n, \tau)$ fulfill case (E2), we have $S_{n, \tau} \in \mc{S}_{n, d_{n}^{\min} - 1} \subseteq \mc{S}_{n, \tau}$ and $c(S_{n, \tau}) = 3n - 6 \leq n \lceil \log(n) \rceil - 2$.
Else if $(n, \tau)$ fulfill case (E4), we have $S_{n, \tau} \in \mc{S}_{n, \tau}$ and $c(S_{n, \tau}) = n\log(n-1) \leq n \lceil \log(n) \rceil - 2$.
Otherwise, $(n, \tau)$ fulfill case (E5) and we have $n \geq 4$ and $\tau \geq \lceil \log(n) \rceil$.
We continue the proof for this case.

For any $(m, n_0, \ldots, n_m, \tau_0, \ldots, \tau_m)$ fulfilling \eqref{eqn: condition}, construct $S$ from $S_{n_i, \tau_i}, \forall i \in [0, m]$ via Construction \ref{construction: small2large}.
According to Lemma \ref{lemma: small2large}, and further noting that $\pi(S) \geq \lceil \log(n) \rceil$, we have $S \in \mc{S}_{n, \tau}$.
Let $(m, n_0, n_1, n_2, \tau_0, \tau_1, \tau_2) = (2, 2, \lceil n/2 \rceil, \lfloor n/2 \rfloor, 0, \tau-1, \tau-1)$.
Since $n \geq 4$ and $\tau \geq \lceil \log(n) \rceil$, for any $i \in [0, 2]$, we have $n_i \in [2, n-1]$ and $\tau_i \geq \lceil \log(n_i - 1) \rceil$.
To continue proof by induction, we assume that for any $i \in [0, 2]$, $S_{n_i, \tau_i}$ exists and fulfills \eqref{eqn: c(S) upper bound}, which must be true for $n_i < 4$ as discussed previously.
We can then easily verify that $(m, n_0, n_1, n_2, \tau_0, \tau_1, \tau_2) = (2, 2, \lceil n/2 \rceil, \lfloor n/2 \rfloor, 0, \tau-1, \tau-1)$ fulfill \eqref{eqn: condition}.
Construct $S$ from $S_{n_i, \tau_i}, \forall i \in [0, 2]$ via Construction \ref{construction: small2large}.
As a result, we have $S \in \mc{S}_{n, \tau}$.
Moreover, according to Lemma \ref{lemma: small2large}, we have $c(S) = c(S_{n_0, \tau_0}) + \sum_{i \in [2]} \big(c(S_{n_i, \tau_i}) + n_i + 1\big) \leq \sum_{i \in [2]} \big(n_i \lceil \log(n_i) \rceil + n_i - 1\big) \leq n \lceil \log(n) \rceil - 2$.
Since $S$ is a candidate for $S_{n, \tau}$, we have $S_{n, \tau} \in \mc{S}_{n, \tau}$ and $c(S_{n, \tau}) \leq n \lceil \log(n) \rceil - 2$.
This completes the proof.


%
%

\ifCLASSOPTIONcaptionsoff
  \newpage
\fi

\bibliographystyle{IEEEtran}
\bibliography{myreference}

\begin{thebibliography}{10}
\providecommand{\url}[1]{#1}
\csname url@samestyle\endcsname
\providecommand{\newblock}{\relax}
\providecommand{\bibinfo}[2]{#2}
\providecommand{\BIBentrySTDinterwordspacing}{\spaceskip=0pt\relax}
\providecommand{\BIBentryALTinterwordstretchfactor}{4}
\providecommand{\BIBentryALTinterwordspacing}{\spaceskip=\fontdimen2\font plus
\BIBentryALTinterwordstretchfactor\fontdimen3\font minus
  \fontdimen4\font\relax}
\providecommand{\BIBforeignlanguage}[2]{{%
\expandafter\ifx\csname l@#1\endcsname\relax
\typeout{** WARNING: IEEEtran.bst: No hyphenation pattern has been}%
\typeout{** loaded for the language `#1'. Using the pattern for}%
\typeout{** the default language instead.}%
\else
\language=\csname l@#1\endcsname
\fi
#2}}
\providecommand{\BIBdecl}{\relax}
\BIBdecl

\bibitem{Gallager62}
R.~G. Gallager, ``Low-density parity-check codes,'' \emph{IRE Trans. Inf.
  Theory}, vol. IT-8, no.~1, pp. 21--28, Jan. 1962.

\bibitem{Richardson01capacity}
T.~J. Richardson and R.~L. Urbanke, ``The capacity of low-density parity-check
  codes under message-passing decoding,'' \emph{IEEE Trans. Inf. Theory},
  vol.~47, no.~2, pp. 599--618, Feb. 2001.

\bibitem{Chen05}
J.~Chen, A.~Dholakia, E.~Eleftheriou, M.~P. Fossorier, and X.-Y. Hu,
  ``Reduced-complexity decoding of {LDPC} codes,'' \emph{IEEE Trans. Commun.},
  vol.~53, no.~8, pp. 1288--1299, Aug. 2005.

\bibitem{he2019onmutual}
X.~He, K.~Cai, and Z.~Mei, ``On mutual information-maximizing quantized belief
  propagation decoding of {LDPC} codes,'' in \emph{Proc. IEEE Global Commun.
  Conf.}, Dec. 2019, pp. 1--6.

\bibitem{he2019mutual}
\BIBentryALTinterwordspacing
------, ``Mutual information-maximizing quantized belief propagation decoding
  of regular {LDPC} codes,'' \emph{arXiv}, 2019. [Online]. Available:
  \url{https://arxiv.org/abs/1904.06666}
\BIBentrySTDinterwordspacing

\bibitem{he2019onfinite}
X.~{He}, K.~{Cai}, and Z.~{Mei}, ``On finite alphabet iterative decoding of
  {LDPC} codes with high-order modulation,'' \emph{IEEE Commun. Lett.},
  vol.~23, no.~11, pp. 1913--1917, Nov. 2019.

\bibitem{hu2001efficient}
X.-Y. Hu, E.~Eleftheriou, D.-M. Arnold, and A.~Dholakia, ``Efficient
  implementations of the sum-product algorithm for decoding {LDPC} codes,'' in
  \emph{Proc. IEEE Global Commun. Conf.}, vol.~2, Nov. 2001, pp. 1036--1036E.

\bibitem{IEEESTD802_11n}
\emph{IEEE standard for information technology---telecommunications and
  information exchange between systems---local and metropolitan area
  networks-specific requirements Part 11: Wireless {LAN} Medium Access Control
  {(MAC)} and Physical Layer {(PHY)} Specifications}, IEEE Std. 802.11n, Oct.
  2009.

\bibitem{wey2008algorithms}
C.-L. Wey, M.-D. Shieh, and S.-Y. Lin, ``Algorithms of finding the first two
  minimum values and their hardware implementation,'' \emph{IEEE Trans.
  Circuits Syst. I: Reg. Papers}, vol.~55, no.~11, pp. 3430--3437, Dec. 2008.

\bibitem{lee2015low}
Y.~Lee, B.~Kim, J.~Jung, and I.-C. Park, ``Low-complexity tree architecture for
  finding the first two minima,'' \emph{IEEE Trans. Circuits Syst. II: Exp.
  Briefs}, vol.~62, no.~1, pp. 61--64, Jan. 2015.

\bibitem{Kurkoski08}
B.~M. Kurkoski, K.~Yamaguchi, and K.~Kobayashi, ``Noise thresholds for discrete
  {LDPC} decoding mappings,'' in \emph{Proc. IEEE Global Commun. Conf.}, Dec.
  2008, pp. 1--5.

\bibitem{Romero15decoding}
F.~J.~C. Romero and B.~M. Kurkoski, ``Decoding {LDPC} codes with mutual
  information-maximizing lookup tables,'' in \emph{Proc. IEEE Int. Symp. Inf.
  Theory}, Jun. 2015, pp. 426--430.

\bibitem{Romero16}
------, ``{LDPC} decoding mappings that maximize mutual information,''
  \emph{IEEE J. Sel. Areas Commun.}, vol.~34, no.~9, pp. 2391--2401, Sep. 2016.

\bibitem{meidlinger2020design}
M.~Meidlinger, G.~Matz, and A.~Burg, ``Design and decoding of irregular {LDPC}
  codes based on discrete message passing,'' \emph{IEEE Trans. Commun.},
  vol.~68, no.~3, pp. 1329--1343, Mar. 2020.

\bibitem{Meidlinger15}
M.~Meidlinger, A.~Balatsoukas-Stimming, A.~Burg, and G.~Matz, ``Quantized
  message passing for {LDPC} codes,'' in \emph{Proc. 49th Asilomar Conf.
  Signals, Syst., Comput.}, Nov. 2015, pp. 1606--1610.

\bibitem{Lewandowsky16}
J.~Lewandowsky, M.~Stark, and G.~Bauch, ``Optimum message mapping {LDPC}
  decoders derived from the sum-product algorithm,'' in \emph{Proc. IEEE Int.
  Conf. Commun.}, May 2016, pp. 1--6.

\bibitem{balatsoukas2015fully}
A.~Balatsoukas-Stimming, M.~Meidlinger, R.~Ghanaatian, G.~Matz, and A.~Burg,
  ``A fully-unrolled {LDPC} decoder based on quantized message passing,'' in
  \emph{Proc. IEEE Workshop on Signal Processing Systems}, Oct. 2015, pp. 1--6.

\bibitem{ghanaatian2018a588}
R.~Ghanaatian, A.~Balatsoukas-Stimming, T.~C. M{\"u}ller, M.~Meidlinger,
  G.~Matz, A.~Teman, and A.~Burg, ``A 588-gb/s {LDPC} decoder based on
  finite-alphabet message passing,'' \emph{IEEE Trans. on Very Large Scale
  Integration (VLSI) Systems}, vol.~26, no.~2, pp. 329--340, Feb. 2018.

\bibitem{Lewandowsky18}
J.~Lewandowsky and G.~Bauch, ``Information-optimum {LDPC} decoders based on the
  information bottleneck method,'' \emph{IEEE Access}, vol.~6, pp. 4054--4071,
  Jan. 2018.

\bibitem{lewandowsky2018design}
J.~Lewandowsky, G.~Bauch, M.~Tschauner, and P.~Oppermann, ``Design and
  evaluation of information bottleneck {LDPC} decoders for software defined
  radios,'' in \emph{Proc. Int. Conf. Signal Processing and Commun. Systems},
  Dec. 2018, pp. 1--9.

\bibitem{lewandowsky2019design}
------, ``Design and evaluation of information bottleneck {LDPC} decoders for
  digital signal processors,'' \emph{IEICE Trans. Commun.}, vol. 102, no.~8,
  pp. 1363--1370, Aug. 2019.

\bibitem{cormen2009introduction}
T.~H. Cormen, C.~E. Leiserson, R.~L. Rivest, and C.~Stein, \emph{Introduction
  to Algorithms: 3rd Edition}.\hskip 1em plus 0.5em minus 0.4em\relax
  Cambridge, MA, USA: MIT Press, 2009.

\end{thebibliography}

\begin{IEEEbiographynophoto}{Xuan He}
received the B.E., M.E., and PhD degrees in communication and information systems from the University of Electronic Science and Technology of China (UESTC), Chengdu, China, in 2011, 2013, and 2018, respectively.
From Oct. 2016 to Sep. 2017, he was a Visiting Student sponsored by the China Scholarship Council (CSC) with the University of Waterloo, Waterloo, ON, Canada.
From Oct. 2018 to Dec. 2020, he was a Postdoctoral Research Fellow with the Singapore University of Technology and Design (SUTD), Singapore.
He is now with the Southwest Jiaotong University (SWJTU), Chengdu, China.
His main research interests include coding theory and information theory.
\end{IEEEbiographynophoto}

\begin{IEEEbiographynophoto}{Kui Cai}
received B.E. degree in information and control engineering from Shanghai Jiao Tong University, Shanghai, China, M.Eng degree in electrical engineering from National University of Singapore, and joint Ph.D. degree in electrical engineering from Technical University of Eindhoven, The Netherlands, and National University of Singapore. Currently, she is an Associate Professor with Singapore University of Technology and Design (SUTD). Cai Kui is a senior member of IEEE. She received 2008 IEEE Communications Society Best Paper Award in Coding and Signal Processing for Data Storage. She served as the Vice-Chair (Academia) of IEEE Communications Society, Data Storage Technical Committee (DSTC) during 2015 and 2016. Her main research interests are in the areas of coding theory, information theory, and signal processing for various data storage systems and digital communications.
\end{IEEEbiographynophoto}

\begin{IEEEbiographynophoto}{Liang Zhou}
received the B.E. and M.E. degrees from the University of Electronic Science and Technology of China (UESTC), Chengdu, China, in 1982 and 1984, respectively. He is currently a Professor with the National Key Laboratory of Science and Technology on Communications, UESTC, and the Center for Cyber Security, UESTC. His research interests include the error control coding and coded modulation, the pseudorandom sequence, the secure communication, and the cryptography.
\end{IEEEbiographynophoto}

\end{document}